\input harvmac.tex

\lref\Witten{E.~Witten, ``Non-Perturbative Superpotentials In
String Theory", Nucl.Phys. {\bf B474} (1996) 343.}
\lref\Mayr{P.~Mayr, ``Mirror Symmetry, N=1 Superpotentials and
Tensionless Strings on Calabi-Yau Four-Folds",
Nucl.Phys. {\bf B494} (1997) 489.}
\lref\Klemm{A.~Klemm, B.~Lian, S.-S.~Roan, S.-T.~Yau,
``Calabi-Yau fourfolds for M- and F-Theory compactifications",
Nucl.Phys. {\bf B518} (1998) 515.}
\lref\Wflux{E. Witten, ``On Flux Quantization In M-Theory And
The Effective Action", J. Geom. Phys. {\bf 22} (1997) 1.}
\lref\GVW{S.~Gukov, C.~Vafa and E.~Witten, ``CFT's From
Calabi-Yau Four-folds", hep-th/9906070.}
\lref\BB{K.~Becker, M.~Becker, ``M-Theory on Eight-Manifolds",
Nucl.Phys. {\bf B477} (1996) 155.}
\lref\SVW{S. Sethi, C. Vafa, and E. Witten,
``Constraints on Low-Dimensional String Compacti\-fi\-cations",
Nucl. Phys. {\bf B480} (1996) 213.}
\lref\HL{M.~Haack, J.~Louis, ``Duality in Heterotic Vacua
With Four Supercharges", hep-th/9912181.}
\lref\COssa{P.~Candelas, X.~C.~de la Ossa, ``Moduli Space of
Calabi-Yau Manifolds", Nucl.Phys. {\bf B355} (1991) 455.}
\lref\Lerche{W.~Lerche, ``Fayet-Iliopoulos Potentials from
Four-Folds", JHEP {\bf 9711} (1997) 004.}
\lref\Vafa{S.~Katz, P.~Mayr and C.~Vafa, ``Mirror
symmetry and Exact Solution of 4D $N=2$ Gauge Theories -- I",
Adv.Theor.Math.Phys. {\bf 1} (1998) 53.}
\lref\ftheory{C.~Vafa, ``Evidence for F-Theory", Nucl. Phys. {\bf B496}
(1996) 403.}
\lref\DM{K.~Dasgupta, S.~Mukhi, ``A Note on Low-Dimensional String
Compactifications", Phys. Lett. {\bf B398} (1997) 285.}
\lref\Gukov{S.~Gukov, ``Solitons, Superpotentials and Calibrations",
hep-th/9911011.}
\lref\LVW{W.~Lerche, C.~Vafa and N.P.~Warner, ``Chiral Rings in
N=2 Superconformal Theories", Nucl.Phys. {\bf B324} (1989) 427.}
\lref\Dixon{L.~Dixon, Lectures at the 1987 ICTP summer Workshop
in High Energy Physics and Cosmology.}
\lref\mirrorbook{S.-T.~Yau, editor, {\it Essays on Mirror Manifolds},
International Press, 1992; B.~Greene and S.-T.~Yau, editors,
{\it Mirror Symmetry. II}, International Press, 1997.}
\lref\Klemmlect{A.~Klemm, ``On the Geometry behind N=2
Supersymmetric Effective Actions in Four Dimensions", hep-th/9705131.}

\lref\HP{P.S.~Howe and G.~Papadopoulos, ``N=2, D = 2 Supergeometry",
Class. Quantum Grav. {\bf4} (1987) 11.}
\lref\GW{M.T.~Grisaru and M.E.~Wehlau, ``Prepotentials for (2,2)
Supergravity", Int. J. Mod. Phys. {\bf A10} (1995) 753.}
\lref\GWmeasure{M.T.~Grisaru, M.E.~Wehlau, ``Superspace Measures,
Invariant Actions, and Component Projection Formulae for $(2,2)$
Supergravity", Nucl. Phys. {\bf B457} (1995) 219.}
\lref\GGW{S.J.~Gates, Jr., M.T.~Grisaru and M.E.~Wehlau,
``A Study of General 2D, N=2 Matter Coupled to Supergravity in
Superspace", Nucl.Phys. {\bf B460} (1996) 579.}
\lref\G{S.J.~Gates, ``Ectoplasm Has No Topology",
Nucl.Phys. {\bf B541} (1999) 615.}
\lref\A{A.~Alnowaiser, ``Supergravity with N=2 in
Two Dimensions", Class. Quantum Grav. {\bf 7} (1990) 1033.}
\lref\IKR{I.T.~Ivanov, B.~Kim and M.~Rocek,
``Complex Structures, Duality, and WZW-Models in Extended
Superspace", Phys.Lett. {\bf 343} (1995) 133.}
\lref\BLR{T.~Buscher, U.~Lindstrom, M.~Rocek, ``New Supersymmetric
Sigma Models with Wess-Zumino Terms", Phys. Lett. {\bf B202} (1988) 94.}
\lref\ST{A.~Sevrin and J.~Troost, ``Off-Shell Formulation of
N=2 Non-Linear Sigma-Models", Nucl.Phys. {\bf B492} (1997) 623.}
\lref\GGRS{S.J.~Gates, M.T.~Grisaru, M.~Rocek and W.~Siegel,
``Superspace'', Benjamin-Cummings, 1983.}
\lref\GHR{S.J.~Gates, C.M.~Hull and M.~Rocek,
``Twisted Multiplets and New Supersymmetric Nonlinear Sigma Models",
Nucl.Phys. {\bf B248} (1984) 157.}
\lref\AGCG{L.~Alvarez-Gaume, S.~Coleman, P.~Ginsparg,
``Finiteness of Ricci Flat N=2 Super\-sym\-metric Sigma Models",
Commun. Math. Phys. {\bf 103} (1986) 423.}
\lref\Gates{S.~J.~Gates, Jr., ``Vector Multiplets and
the Phases of N = 2 Theories in 2D Through the Looking Glass",
Phys. Lett. {\bf B352} (1995) 43.}
\lref\GKNO{S.J.~ Gates, T.~Kadoyoshi, S.~Nojiri,
S.D.~Odintsov, ``Quantum cosmology in the models of
2d and 4d dilatonic supergravity with WZ matter",
Phys.Rev. {\bf D58} (1998) 084026.}
\lref\WGRN{B.~de~Wit, M.T.~Grisaru, E.~Rabinovici, H.~Nicolai,
``Two-loop finiteness of D=2 supergravity",
Phys.Lett. {\bf B286} (1992) 78.}
\lref\Nishino{H.~Nishino, ``Alternative $N=(4,0)$ Superstring
and $\sigma$-Models", Phys.Lett. {\bf B355} (1995) 117.}
\lref\GatesRana{S.~J.~Gates, Jr., L.~Rana, ``Manifest $(4,0)$
Supersymmetry, Sigma Models and The ADHM Instaton Construction",
Phys. Lett. {\bf B345} (1995) 233.}
\lref\DGR{R.~Dhanawittayapol, S.~J.~Gates Jr., L.~Rana,
``A Canticle on (4,0) Supergravity-Scalar Multiplet Systems for
a {\it Cognoscente}", Phys.Lett. {\bf B389} (1996) 264.}
\lref\BS{E.~Bergshoeff, E.~Sezgin, ``The $(4,0)$ Heterotic String
with Wess-Zumino Term", Mod. Phys. Lett. {\bf A1} (1986) 191.}
\lref\WADHM{E.~Witten, ``Sigma Models And The ADHM Construction
Of Instantons", J. Geom. Phys. {\bf 15} (1995) 215.}
\lref\HPT{C.~M.~Hull, G.~Papadopoulos, P.~K.~Townsend,
``Potentials for $(p,0)$ and $(1,1)$ Supersymmetric Sigma Models
with Torsion", Phys.Lett. {\bf B316} (1993) 291.}
\lref\GS{A.~Galperin, E.~Sokatchev, ``Manifest Supersymmetry and the
ADHM Construction of Instantons", Nucl. Phys. {\bf B452} (1995) 431.}
\lref\COvrut{G.~L.~Cardoso and B.~A.~Ovrut,``A Green-Schwarz mechanism
for D = 4, N=1 supergravity anomalies", Nucl. Phys. {\bf B369} (1992) 351.}
\lref\ABGG{P.~Adamietz, P.~Binetruy, G.~Girardi and R.~Grimm,
``Supergravity and matter: Linear multiplet couplings and
Kahler anomaly cancellation,'' Nucl. Phys. {\bf B401} (1993) 257.}
\lref\DFKZ{J.~P.~Derendinger, S.~Ferrara, C.~Kounnas and F.~Zwirner,
``On loop corrections to string effective field theories: Field dependent
gauge couplings and sigma model anomalies,''
Nucl. Phys. {\bf B372} (1992) 145.}
\lref\CGHS{C.~Callan, S.~Giddings, J.~Harvey, A.~Strominger,
``Evanescent Black Holes", Phys.Rev. {\bf D45} (1992) 1005.}

\font\cmss=cmss10 \font\cmsss=cmss10 at 7pt

\def\IB{\relax\hbox{$\inbar\kern-.3em{\rm B}$}}
\def\IC{\relax\hbox{$\inbar\kern-.3em{\rm C}$}}
\def\ID{\relax\hbox{$\inbar\kern-.3em{\rm D}$}}
\def\IE{\relax\hbox{$\inbar\kern-.3em{\rm E}$}}
\def\IF{\relax\hbox{$\inbar\kern-.3em{\rm F}$}}
\def\IG{\relax\hbox{$\inbar\kern-.3em{\rm G}$}}
\def\IGa{\relax\hbox{${\rm I}\kern-.18em\Gamma$}}
\def\IH{\relax{\rm I\kern-.18em H}}
\def\IK{\relax{\rm I\kern-.18em K}}
\def\IL{\relax{\rm I\kern-.18em L}}
\def\IP{\relax{\rm I\kern-.18em P}}
\def\IR{\relax{\rm I\kern-.18em R}}
\def\Z{\relax\ifmmode\mathchoice
{\hbox{\cmss Z\kern-.4em Z}}{\hbox{\cmss Z\kern-.4em Z}}
{\lower.9pt\hbox{\cmsss Z\kern-.4em Z}}
{\lower1.2pt\hbox{\cmsss Z\kern-.4em Z}}\else{\cmss Z\kern-.4em Z}\fi}

\def\II{\relax{\rm I\kern-.18em I}}

\def\S{{\bf S}}

\def\CA {{\cal A}}

\def\CE {{\cal E}}
\def\CF {{\cal F}}
\def\CG {{\cal G}}

\def\CK {{\cal K}}
\def\CL {{\cal L}}
\def\CM {{\cal M}}
\def\CN {{\cal N}}

\def\CR {{\cal R}}

\def\CV {{\cal V}}

\def\CX {{\cal X}}
\def\CY {{\cal Y}}
\def\CZ {{\cal Z}}


\def\p{\partial}

\def\D{{\slash\!\!\!\! D}}
\def\DD{{\bf D}}

\def\tilde{\widetilde}
\def\hat{\widehat}
\def\bar{\overline}

\def\nb {{\bar{\nabla}}}


\def\Tr{{\rm Tr}}

\def\p{\partial}

\def\inbar{\,\vrule height1.5ex width.4pt depth0pt}
\def\r{{\rm Re}}
\def\i{{\rm Im}}

\def\a{\alpha}
\def\b{\beta}

\def\d{\delta}

\def\la{\lambda}
\def\th{\theta}
\def\s{\sigma}

\def\p{\partial}
\def\S{\Sigma}

\def\pp{{\mathchoice
              %
          {
              \kern 1pt%
              \raise 1pt
              \vbox{\hrule width5pt height0.4pt depth0pt
                    \kern -2pt
                    \hbox{\kern 2.3pt
                          \vrule width0.4pt height6pt depth0pt
                          }
                    \kern -2pt
                    \hrule width5pt height0.4pt depth0pt}%
                    \kern 1pt
           }
            {
              \kern 1pt%
              \raise 1pt
              \vbox{\hrule width4.3pt height0.4pt depth0pt
                    \kern -1.8pt
                    \hbox{\kern 1.95pt
                          \vrule width0.4pt height5.4pt depth0pt
                          }
                    \kern -1.8pt
                    \hrule width4.3pt height0.4pt depth0pt}%
                    \kern 1pt
            }
            {
              \kern 0.5pt%
              \raise 1pt
              \vbox{\hrule width4.0pt height0.3pt depth0pt
                    \kern -1.9pt  
                    \hbox{\kern 1.85pt
                          \vrule width0.3pt height5.7pt depth0pt
                          }
                    \kern -1.9pt
                    \hrule width4.0pt height0.3pt depth0pt}%
                    \kern 0.5pt
            }
            {
              \kern 0.5pt%
              \raise 1pt
              \vbox{\hrule width3.6pt height0.3pt depth0pt
                    \kern -1.5pt
                    \hbox{\kern 1.65pt
                          \vrule width0.3pt height4.5pt depth0pt
                          }
                    \kern -1.5pt
                    \hrule width3.6pt height0.3pt depth0pt}%
                    \kern 0.5pt
            }
        }}

\def\mm{{\mathchoice
   %
                  %
                       {
                             \kern 1pt
               \raise 1pt    \vbox{\hrule width5pt height0.4pt depth0pt
                                  \kern 2pt
                                  \hrule width5pt height0.4pt depth0pt}
                             \kern 1pt}
                       {
                            \kern 1pt
               \raise 1pt \vbox{\hrule width4.3pt height0.4pt depth0pt
                                  \kern 1.8pt
                                  \hrule width4.3pt height0.4pt depth0pt}
                             \kern 1pt}
                       {
                            \kern 0.5pt
               \raise 1pt
                            \vbox{\hrule width4.0pt height0.3pt depth0pt
                                  \kern 1.9pt
                                  \hrule width4.0pt height0.3pt depth0pt}
                            \kern 1pt}
                       {
                           \kern 0.5pt
             \raise 1pt  \vbox{\hrule width3.6pt height0.3pt depth0pt
                                  \kern 1.5pt
                                  \hrule width3.6pt height0.3pt depth0pt}
                           \kern 0.5pt}
                       }}

\def\ad{{\kern0.5pt
                   \alpha \kern-5.05pt
\raise5.8pt\hbox{$\textstyle.$}\kern 0.5pt}}

\def\bd{{\kern0.5pt
                   \beta \kern-5.05pt \raise5.8pt\hbox{$\textstyle.$}\kern 0.5pt}}

\def\qd{{\kern0.5pt
                   q \kern-5.05pt \raise5.8pt\hbox{$\textstyle.$}\kern 0.5pt}}
\def\Dot#1{{\kern0.5pt
     {#1} \kern-5.05pt \raise5.8pt\hbox{$\textstyle.$}\kern 0.5pt}}

\Title{ \vbox{\baselineskip12pt
\hbox{CALT-68-2258}
\hbox{CITUSC/00-005}
\hbox{PUPT-1904}
\hbox{UMDEPP 00-046}}}
{{\vbox{\centerline{Two Two-Dimensional Supergravity Theories}
\centerline{from Calabi-Yau Four-Folds}}}}
\centerline{S. James Gates, Jr.,
$^{\spadesuit}$\foot{E-Mail: gatess@wam.umd.edu}
Sergei Gukov,$^{\clubsuit \diamondsuit}
$\foot{E-Mail: gukov@feynman.princeton.edu}
and Edward Witten$^{\clubsuit \heartsuit}$}
\bigskip
\bigskip
\centerline{\it $^{\spadesuit}$ Department of Physics,
University of Maryland at College Park,}
\centerline{\it College Park, MD 20742-4111, USA}
\smallskip
\centerline{\it $^{\clubsuit}$ California Institute of Technology,
Pasadena, CA 91125, USA,}
\centerline{\it CIT-USC Center For Theoretical Physics}
\smallskip
\centerline{\it $^{\diamondsuit}$ Joseph Henry Laboratories, Princeton
University,}
\centerline{\it Princeton, NJ 08544, USA}
\smallskip
\centerline{\it $^{\heartsuit}$ School of Natural Sciences, Institute for
Advanced Study,}
\centerline{\it Olden Lane, Princeton, NJ 08540, USA}
\vskip .3in

We consider two-dimensional supergravity theories with
four supercharges constructed from compactification
of Type II string theory on a generic Calabi-Yau four-fold.
In Type IIA and Type IIB cases, respectively,
new superspace formulations of $\CN=(2,2)$ and $\CN=(0,4)$
dilaton supergravities are found and their coupling
to matter multiplets is discussed.

\Date{May 2000}


\newsec{Introduction}

For a long time, compactification of heterotic string
theory on Calabi-Yau manifolds was the primary candidate
for constructing realistic models in four dimensions
with $\CN=1$ supersymmetry. This was also a strong
motivation to study Type II superstrings on Calabi-Yau
three-folds which share many common properties with the
corresponding heterotic compactifications.  At the same
time substantial progress has been made in understanding
the mathematical aspects of Calabi-Yau three-folds, such
as quantum cohomology and mirror symmetry \mirrorbook.
It was not until the discovery of F-theory \ftheory\ that
it was realized that $\CN=1$ four-dimensional
heterotic string vacua can be
equivalently described as F-theory compactifications on
elliptically fibered Calabi-Yau four-folds. Since then,
the study of Calabi-Yau four-fold compactifications has
become of particular importance for physical applications.
Compactification of F-theory on an elliptically fibered
Calabi-Yau four-fold is closely related to the corresponding
compactifications of Type IIA and M-theory.  Namely, when
the area of the elliptic fiber shrinks to zero, M-theory
compactification on a Calabi-Yau four-fold is well described
by F-theory compactification on the same Calabi-Yau manifold.
On the other hand, Type IIA string theory is related to
M-theory via compactification on an extra circle (of small
radius).

One of the most striking outcomes in the study of
compactifications on Calabi-Yau three-folds is the great
success in understanding non-perturbative phenomena in
$\CN=2$ field theories in four dimensions,
see {\it e.g.} \Klemmlect\ for introduction and references.
On the other hand, understanding
of Calabi-Yau four-fold compactifications still is quite
far from that stage, so we shall not discuss non-perturbative
phenomena in this paper.  Instead, we consider classical
supergravity theories interacting with two-dimensional
non-linear sigma-models that can be constructed from
Calabi-Yau four-folds.  Surprisingly, it turns out that
manifestly supersymmetric formulations of such theories has
not been given previously.

On general grounds, a compactification of Type IIA (IIB)
string theory on a Calabi-Yau four-fold leads to a $\CN=(2,2)$
(resp. $\CN=(0,4)$) effective field theory in two dimensions.
In the low-energy limit the theory is described by supergravity
coupled to matter. For example, from the Kaluza-Klein reduction
of Type IIA string theory on a Calabi-Yau four-fold in section
3 we find that a suitable low-energy theory is $\CN=(2,2)$
dilaton supergravity interacting with some number of chiral
and twisted chiral multiplets.  It is invariant
under the ``mirror transformation" which, acting on the matter
fields, exchanges chiral multiplets and twisted chiral multiplets.
We thus generalize the proposal of \Gates\ where it was suggested
that the ``kinematic structure" of the mirror transformation
has its origin in a mapping between chiral and twisted
chiral multiplets when these superfields are regarded as the
fundamental degrees of $\CN = (2,2)$ superstring theories.
The generalization posits that this mapping also applies to
the effective action. Another characteristic feature of this
supergravity theory is that the supergravity multiplet
contains a real dilaton field.
In a special case, when all matter multiplets are chiral and massless,
a component action of this $\CN=2$ dilaton supergravity was
constructed in \WGRN.  However, to describe Type IIA compactifications
on Calabi-Yau four-folds we need a generalization of this theory
that includes interaction with twisted chiral multiplets and the
possibility to turn on the superpotential and as well the twisted
superpotential. Thus, in sections 4 and 5 we present superspace
construction of general $\CN=(2,2)$ dilaton supergravity coupled to matter.
The construction in section 4 is based on the Goldstone mechanism in the
superspace formulation of non-minimal gauged $\CN=(2,2)$ supergravity.
Coupling of the new $\CN=(2,2)$ dilaton supergravity to matter
multiplets is the subject of section 5, where we discuss local
integration in superspace.
In section 6 we perform the Kaluza-Klein reduction of Type IIB
string theory on a Calabi-Yau four-fold and describe the component
action of the resulting $\CN=(0,4)$ dilaton supergravity.
Most of this section, as well as section 3, is not new
and presented for the sake of completeness.
The superspace formulation of the new $\CN=(0,4)$ dilaton
supergravity is presented in section 7. In the appendix A
we present a straightforward but technical world-sheet calculation
of string amplitudes corresponding to the target space metric of
the effective two-dimensional theory, and in the appendix B we list
extra derivative constraints arising from de-gauging $\CN=(2,2)$
non-minimal supergravity.
Appendix C contains components of the covariant derivative
in $\CN=(2,2)$ dilaton supergravity needed in section 5.
Finally, in appendix D we repeat the derivation \GWmeasure\ of
the chiral density projection formula in $\CN=(2,2)$ dilaton
supergravity.
We begin in the next section with a summary of notations and
definitions used throughout the paper.

\newsec{Calabi-Yau Four-folds: Some Conventions and Definitions}

We study compactification of Type II string theory on $M_{(2)}
\times X$ where $M_{(2)}$ is a maximally symmetric homogeneous
two-dimensional space-time and $X$ is a Calabi-Yau four-fold.
We use the following notations for the space-time indices.
Capital letters $M$, $N$, $\ldots$ run from 0 to 9 and denote
ten-dimensional Lorentz indices. Latin letters $m$, $n$, $\ldots$
and $a$, $b$, $\ldots$ represent, respectively, real and holomorphic
indices tangent to $X$. Greek letters $\a$, $\b$, $\ldots$ and $\mu$,
$\nu$, $\ldots$ are used for the two-dimensional spinor indices
``$+$" and ``$-$", and light-cone indices ``$\pp$" and ``$\mm$",
correspondingly. Sometimes we also use capital latin letters
$A$, $B$, $\ldots$ to denote both spinor and vector indices.

A Calabi-Yau space $X$ is a compact K\"ahler manifold with complex
dimension four and $SU(4)$ holonomy group.  It follows that $X$ is
a Ricci-flat manifold and, therefore, it can be used as a background
for Type II string compactification.  As a topological space, $X$
is classified by the Hodge numbers $h^{p,q}$ which count the number
of harmonic $(p,q)$-forms $\omega_i^{(p,q)} \in H^{p,q} (X)$, $i =
1, \ldots h^{p,q}$. The non-vanishing cohomology groups have the
following dimensions \Klemm:
$$
h^{1,1} = h^{3,3}, \quad \quad h^{3,1} = h^{1,3},
$$
$$
h^{2,1} = h^{1,2} = h^{3,2} = h^{2,3},
$$
$$
h^{0,0} = h^{4,4} = h^{4,0} = h^{0,4} = 1,
$$
\eqn\htwotwo{h^{2,2} =  2(22 + 2 h^{1,1} + 2 h^{3,1} - h^{2,1}).}
For the Euler number of $X$ we have:
\eqn\euler{ {\chi \over 6} = 8 + h^{1,1} + h^{3,1} - h^{2,1}}

We denote by $\Omega$ a covariantly constant $(4,0)$-form.  The
$(1,1)$- and $(3,1)$-forms are related to the deformation parameters
of the K\"ahler form and the complex structure of $X$, respectively.
Namely, an arbitrary variation of the metric of the Calabi-Yau
four-fold $X$ that respects $SU(4)$ holonomy looks like:
$$
\delta g_{a \bar b} dz^a d \bar z^{\bar b}
+ \delta g_{ab} dz^a dz^b + {\rm c.c.}
$$
where
\eqn\defs{\delta g_{a b} =
\sum_{j=1}^{h^{3,1}} \phi^j w^j_{a b},
\quad \quad i \delta g_{a \bar b} =
\sum_{i=1}^{h^{1,1}} s^i \omega^i_{a \bar b}.}
By the appropriate contraction with $\bar \Omega$, from the
forms $w^j_{ab}$ we can construct elements in $H^{1,3} (X)$:
$$
\omega_j^{(1,3)} = \bar \Omega_{\bar a \bar b \bar c \bar d}
g^{\bar d d} w^j_{d f} d \bar z^{\bar a} d \bar z^{\bar b}
d \bar z^{\bar c} d z^{f}
$$

In what follows we will use some integrals over the Calabi-Yau
space $X$ \refs{\COssa,\HL}:
$$
\CV = \int_X d^8z \sqrt{g} =
{1 \over 4!} \int_X \CK \wedge \CK \wedge \CK \wedge \CK
$$
\eqn\phimet{ \CG_{\phi_i \bar \phi_{\bar j}} =
{1 \over 4 \CV} \int_X d^8z \sqrt{g}~w_{i~ab} \bar w_{\bar j}^{ab} }
\eqn\smet{\CG_{\s_k \bar \s_l} = {1 \over 2 \CV} \int_X d^8z \sqrt{g}~
\omega^k_{a \bar b} \omega^{l~a \bar b} }
\eqn\ytri{Y_{i m \bar n} = \int_X \omega^{(1,1)}_i \wedge
\omega^{(2,1)}_m \wedge \omega^{(1,2)}_{\bar n}}
\eqn\dijkl{ d_{ijkl} = \int_X \omega^{(1,1)}_i \wedge
\omega^{(1,1)}_j \wedge \omega^{(1,1)}_k \wedge \omega^{(1,1)}_l}
The notations \phimet\ and \smet\ will become clear in the next
section where we identify expectation values of the fields $s_i$ and
$\phi_j$ with the K\"ahler and complex structure moduli, respectively.
In particular, we write:
\eqn\jdef{\CK = i g_{a \bar b} dz^a \wedge d \bar z^{\bar b} =
\sum_{i=1}^{h^{1,1}} \langle s_i\rangle~ \omega_i^{(1,1)}}
for the K\"ahler form on $X$.
The moduli space of a Calabi-Yau space is locally a product of
the moduli space of complex deformations, $\CM_c (X)$, and the
(complexified) moduli space of K\"ahler structure, $\CM_{\CK} (X)$.
Notice, the metric $\CG_{\phi_i \bar \phi_{\bar j}}$ defined above
is the Weil-Petersson metric on the moduli space of complex
structure of $X$, with the K\"ahler potential, {\it cf.} \COssa:
\eqn\kphi{K (\phi_i, \bar \phi_{\bar i}) =
- \ln \Big( \int_X \Omega \wedge \bar \Omega \Big).}

\newsec{Compactification of Type IIA String Theory on Calabi-Yau
Four-folds}

In this section we describe the effective two-dimensional theory
constructed from compactification of Type IIA string theory on a
Calabi-Yau four-fold $X$.  When the volume of $X$ is large compared
to the string scale, Type IIA supergravity is a good low-energy
approximation to Type IIA string theory.  Therefore, in the `large
volume limit' we may describe the low-energy effective theory studying
compactification of Type IIA supergravity on the Calabi-Yau space $X$.
With this motivation, let us start this section recalling some
facts about Type IIA supergravity itself.

The bosonic field content of Type IIA supergravity contains the
metric $g_{MN}$, the dilaton $\varphi$, a vector field $A_M$,
and tensor fields $B_{MN}$ and $C_{MNP}$.
The bosonic part of the Lagrangian (in string frame) looks like:
\eqn\alagr{ L_{(10)} = \sqrt{-g} \Big[
\half e^{-2 \varphi} ( R^{(10)} + 4 \big(\nabla \varphi)^2
- {1 \over 12} H^2 \big)
- {1 \over 4} F^2 -  {1 \over 48} G'^2  \Big] + \ldots }
where we introduced the gauge-invariant field strengths:
$$
F = dA, \quad \quad H = d B,
$$
$$
G = dC, \quad \quad G' = G + A \wedge H.
$$
With this choice of normalization the fields $A$, $B$,
and $C$ transform in a natural, dilaton-independent way
under gauge transformations.

The dots in the Lagrangian \alagr\ stand for higher order terms
among which we find the Chern-Simons term $B \wedge G \wedge G$
and the anomaly term $B \wedge I_8$, where the eight-form $I_8$
is proportional to the Euler density of $X$. After integration
over a compact eight-manifold $X$ these topological terms produce
a global anomaly \refs{\SVW,\Wflux}:
\eqn\sethrel{N={\chi\over 24}-{1\over 2 (2\pi)^2}\int_X G \wedge G}
To cancel the tadpole for the $B$-field one has to introduce $N$
fundamental strings filling two-dimensional non-compact space.

The action of Type IIA supergravity is invariant under 16 left
and 16 right supersymmetry transformations, such that the left
supersymmetries are chiral while the right supersymmetries are
anti-chiral with respect to the ten-dimensional chirality operator
$\Gamma_{11}$. Since $X$ admits a nowhere vanishing complex spinor
of definite chirality, compactification of Type IIA string theory
on $X$ is described by $\CN=(2,2)$ supergravity theory coupled to
matter.  With the appropriate choice of orientation, the fundamental
strings filling two-dimensional space-time do not break supersymmetry
further.

To find the spectrum of the effective low-energy theory we perform
Kaluza-Klein reduction of Type IIA supergravity to two dimensions.
Below we describe the decomposition of Type IIA bosonic fields in
harmonics of $X$. By supersymmetry, incorporation of fermionic
zero-modes completes the resulting spectrum into appropriate $\CN=(2,2)$
supermultiplets. Via dimensional reduction Type IIA dilaton $\varphi$
becomes a real scalar field in the two-dimensional theory.
The ten-dimensional metric $g_{MN}$ decomposes into
the two-dimensional metric
$g_{\mu \nu}$,
$h^{3,1}$ complex scalars $\phi_i$ and
$h^{1,1}$ real scalars $s_j$ defined in \defs.
The antisymmetric tensor fields $B_{MN}$ and $C_{MNP}$
can be expanded into harmonic modes as follows:
\eqn\bmodes{B = \sum_{i=1}^{h^{1,1}} r^i \omega_i^{(1,1)},}
\eqn\cmodes{C = \sum_{j=1}^{h^{1,1}} A^j_{\mu} \omega_j^{(1,1)}
+ \sum_{k=1}^{h^{2,1}} z^k \omega_k^{(2,1)} + {\rm c.c.}}
It is convenient to combine real fields $s^i$ and $r^i$ into
complex scalars $\sigma^i$.
Taking into account the vector field $A_{\mu}$ from
the Ramond-Ramond sector of Type IIA theory, we end up
with the following list of $\CN=(2,2)$ supermultiplets:
$$
{\rm a~ gravitational~ multiplet:}
\quad g_{\mu \nu},
\quad A_{\mu},
\quad \varphi
$$
$$
h^{3,1}~ {\rm chiral~ multiplets:}
\quad \phi_i,
\quad \bar \phi_{\bar i}
$$
$$
h^{1,1}~ {\rm twisted~ chiral~ multiplets:}
\quad \sigma^j,
\quad \bar \sigma^j,
\quad A^j_{\mu}
$$
$$
h^{2,1}~ {\rm (twisted)~chiral~ multiplets:}
\quad z^k, \quad \bar z^{\bar k}
$$
Vector fields $A_{\mu}$ and $A^i_{\mu}$ do not have propagating
degrees of freedom in two dimensions and play the role of auxiliary
fields in the supergravity multiplet and twisted chiral multiplets,
respectively.

The complex scalar fields $z^k$ which come from $(2,1)$-modes
take value in a torus. When background fluxes satisfy
$G \wedge \omega_k^{(2,1)} =0$ and $H \wedge ^* \omega_k^{(2,1)} =0$
there is no superpotential for the corresponding harmonics $z_k$,
so these fields are massless. A $T$-duality transformation on the
torus then converts them from ordinary chiral superfields
to twisted chiral superfields. It is natural to choose $z_k$
to be scalar  components of chiral superfields.
Indeed, if we start in eleven dimensions, the reduction of
the $C$-field \cmodes\ yields $h^{2,1}$ complex scalar modes
in three dimensions.
These modes are scalar components of chiral superfields
since there is no notion of ``twisted chiral superfields"
in three dimensions. After a further compactification on
a circle they naturally remain as chiral superfields in
two dimensions, but now a $T$-duality becomes possible
and $z_k$ can be alternatively described if one prefers
as twisted chiral superfields.

To find the effective action for the light fields we have
to substitute \defs, \bmodes\ and \cmodes\ in
the Lagrangian \alagr\ and integrate over the internal space $X$.
Using the formulas \phimet, \smet\ and \ytri\ we obtain
the following effective action for the bosonic modes,
{\it cf.} \refs{\HL,\WGRN}:
\eqn\act{ L_{(2)} = e^{- 2 \varphi} \CV
\big[ R^{(2)} + 4 (\nabla \varphi)^2
- \CG_{\phi_i \bar \phi_{\bar j}}
(\p_{\mu} \phi^i) (\p^{\mu} \bar \phi^{\bar j}) -}
$$
- {1 \over 2} \CG_{\s_i \bar \s_j}
(\p_{\mu} \s^i) (\p_{\mu} \bar \s^{j}) \big]
- {1 \over 4} Y_{i m \bar n} \s^i
(D_{\mu} z^m) (D^{\mu} \bar z^{\bar n})
+ \ldots
$$
where the covariant derivative $D_{\mu}$ acting on $z^m$
contains a connection corresponding to the holomorphic dependence
of the basis of $(2,1)$-forms on the complex structure \HL.

To summarize, we find that in the large volume limit compactification
of Type IIA string theory on a Calabi-Yau four-fold $X$ leads to
$\CN=(2,2)$ dilaton supergravity coupled to a non-linear sigma-model.
The target space of this sigma-model is parametrized by some
number of chiral and twisted chiral multiplets.
This agrees with the result of \ST,
where it was found that the most general $\CN=(2,2)$ non-linear
sigma-model is based on a target space with two non-commuting
complex structures $J_{\pm}$, so that the space ${\rm ker} (J_+ - J_-)$
is parametrized by chiral superfields, while ${\rm ker} (J_+ + J_-)$
is parametrized by twisted chiral superfields.

This kind of dilaton supergravity coupled to $\CN=(2,2)$ chiral
matter was studied some time ago \WGRN.   However, for our purposes
we need to generalize the component construction of \WGRN\ to
include twisted chiral multiplets. Furthermore, background
fluxes of Ramond-Ramond field strengths induce effective superpotential
\refs{\GVW,\Gukov} and/or twisted chiral superpotential \refs{\GVW,
\Gukov, \Lerche} in the two-dimensional theory. Hence, we have to
incorporate these terms in the construction as well. The most
elegant and convenient way to do this is in $\CN=(2,2)$ superspace
where the supersymmetry becomes manifest \refs{\GGRS, \HP}.
In addition to the usual space-time coordinates $x^{\mu}$,
$\CN=(2,2)$ superspace is parametrized by anti-commuting coordinates
$\th^{\a} = (\th^{+}, \th^{-})$ and their complex conjugates
$\bar \th^{\dot \a} = (\bar \th^{\dot +}, \bar \th^{\dot -})$.
Then, we expect that the action of the matter fields \act\
can be written in a compact form, similar to the action
of matter coupled $\CN=1$ supergravity in four dimensions:
\eqn\acti{S = \int d^2x \int d^2 \th d^2 \bar \th E^{-1} \exp ( -  K).}
We postpone the discussion of the superspace measure $E$ till the
next sections where superspace formulation will be discussed in
detail.  Now we simply assume that the suitable measure exists.
The main advantage of the superspace formulation is that due to the
extended supersymmetry, all the term in the action \act\ with up to
two derivatives or four fermions are determined by a single real
function $K(\phi_{i}, \bar \phi_{\bar i}, \s_{j}, \bar \s_{j},
z_k, \bar z_{\bar k}, \varphi)$, the K\"ahler
potential \GHR. It is invariant under the generalized K\"ahler
transformation:
\eqn\ktransform{ K \longrightarrow K
+ \Lambda_1 (\phi_i, \s_j, z_k)
+ \overline \Lambda_1 (\overline \phi_{\bar i},
\overline \s_j, \bar z_{\bar k})
+ \Lambda_2 (\phi_i, \bar \s_j, \bar z_{\bar k})
+ \overline \Lambda_2 (\bar \phi_{\bar i}, \s_j, z_k)}
The target space metric is given by the second
derivative of the K\"ahler potential.

For the sake of simplicity, let us assume for a moment
that $h^{2,1}=0$. Then the metric is block diagonal:
\eqn\split{ \CG_{\phi_i \bar \s_j} =
{\p^2 K \over \p \phi_i \p \bar \s_j} =0}
{}From the condition \split\ it follows that locally we can write
the K\"ahler potential that gives the effective action \act\ as:
\eqn\ktot{K = K_c(\phi_{i}, \bar \phi_{\bar i})
+ K_{\CK} (\s_{j}, \bar \s_{j}) }
where $K_c$ is the K\"ahler potential \kphi\ on the moduli space
of the complex structure:
$$
K_c (\phi_{i}, \bar \phi_{\bar i}) =
- \ln \Big( \int_X \Omega \wedge \bar \Omega \Big).
$$
Similar to the case of Calabi-Yau three-folds \COssa, one can verify
that the metric $\CG_{\s_i \bar \s_j}$ can be obtained from the K\"ahler
potential:
\eqn\ksig{K_{\CK} (\s_{j}, \bar \s_{j}) =
- \ln \Big( \int_X \CK \wedge \CK \wedge \CK \wedge \CK \Big).}
Indeed, if $\omega^{(2)}$ is a harmonic 2-form on a Calabi-Yau four-fold
$X$, its Hodge dual is given by the following neat formula:
\eqn\twoform{^* \omega^{(2)} = - {1 \over 2} \omega^{(2)}
\wedge \CK \wedge \CK + {2 \over 3} {\Big( \int_X \omega^{(2)}
\wedge \CK \wedge \CK \wedge \CK \Big) \over
\Big( \int_X \CK \wedge \CK \wedge \CK \wedge \CK \Big)}
\CK \wedge \CK \wedge \CK}
Therefore, we can write \smet\ in the following form:
$$
\CG_{\s_k \bar \s_l} =
- {1 \over 2 \CV} \int_X \omega^{(1,1)}_k \wedge ^* \omega^{(1,1)}_l =
$$
$$
= {1 \over 4 \CV} \int_X \omega^{(1,1)}_k \wedge
\omega^{(1,1)}_l \wedge \CK \wedge \CK - {1 \over 72 \CV^2}
\Big( \int_X \omega^{(1,1)}_k \wedge \CK \wedge \CK \wedge \CK \Big)
\Big( \int_X \omega^{(1,1)}_l \wedge \CK \wedge \CK \wedge \CK \Big)
$$
Using the explicit expression \jdef\ for the K\"ahler form $\CK$,
it is easy to see that the above metric indeed follows from the
K\"ahler potential \ksig:
$$
\CG_{\s_k \bar \s_l} = - \half {\p^2 K_{\CK} (\s_{j},
\bar \s_{j}) \over \p \s_k \p \bar \s_l}
$$
Hence, to the leading order the metric on the target space is K\"ahler,
torsionless, and equal to the metric on the moduli space
of the Calabi-Yau space $X$, $\CM_c (X) \times \CM_{\CK} (X)$.

The classical action \act\ is invariant under two $U(1)$ $R$-symmetries.
We will denote their linear combinations as $U(1)_A$ and $U(1)_V$.
The action of these symmetries on the supercharges can be represented as:
\eqn\rqcharge{ Q_{-} \quad \quad \bar Q_{\dot +} }
$$
Q_{+} \quad \quad \bar Q_{\dot -}
$$
where the upper (lower) row is assigned a $U(1)_A$ charge $+1$ ($-1$)
while the right (left) column is assigned a $U(1)_V$ charge $+1$ ($-1$).
These R-symmetries are not symmetries of the string theory -- there is
no way to assign $R$-transformations to massive string modes to preserve them.
Even though we will not explicitly include massive string modes in
the present paper, we will include a superpotential and twisted chiral
superpotential that violate the $R$-symmetries. Even in the absence of
the superpotentials, higher derivative interactions among the massless
fields obtained by integrating out massive string states would be expected
to violate the $R$-symmetry.

The explicit expression for the chiral superpotentials generated
by the most general Ramond-Ramond flux $\CF = \sum (RR~ {\rm fields})$
in terms of the Calabi-Yau moduli was derived in \Gukov:
\eqn\wone
{ W(\phi_i) = {1 \over 2 \pi} \int_X \Omega \wedge G}
and for the twisted chiral superpotential:
\eqn\wtwo{\tilde W (\s_j) = {1 \over 2 \pi} \int_X e^{\CK} \wedge \CF}
The superpotential $W(\phi_i)$ and the twisted superpotential $\tilde
W(\s_j)$ are holomorphic functions of the fields $\phi_i$ and $\s_j$,
respectively. Taking into account the superpotential terms, the action
of the matter fields reads as:
\eqn\action{ S_{(2)} = \int d^2x \int d^2 \th d^2 \bar \th E^{-1}
e^{- K} + \int d^2x \int d^2 \th \CE^{-1} W(\phi_i) + }
$$
+ \int d^2x \int d \th^{+} d \th^{\dot -} \tilde \CE^{-1} \tilde W(\s_j)
+ {\rm c.c.}
$$

Generic values of Ramond-Ramond fluxes completely break the $\CN=(2,2)$
supersymmetry\foot{Investigating the supersymmetry conditions
as in \refs{\GVW,\Gukov,\BB}, one can also show that {\it any}
$H$-field flux breaks all the supersymmetry. A simple way to see this is
to assume, on the contrary, that there exists a supersymmetric vacuum
corresponding to a non-zero $H$-flux and consider a BPS soliton
connecting such a vacuum to the vacuum with zero $H$-flux.
In Type IIA string theory this soliton would correspond to an $NS5$-brane
wrapped over a Poincar\'e dual supersymmetric 5-cycle. However,
there is a contradiction since Calabi-Yau 4-folds
do not have supersymmetric 5-cycles, see {\it e.g.} \Gukov.
Therefore, a non-zero $H$-flux lifts all the supersymmetric vacua. 
It is natural to interpret this in terms of the effective superpotential
$W \sim \int C \wedge ^* H$ for the scalar fields $z_k$.}.
In the two-dimensional theory this effect corresponds
to generation of a superpotential that lifts (part of) supersymmetric
vacua. However, if the vacuum values of the fields $\phi_i$ and
$\s_j$ satisfy:
\eqn\minimum{ {D W \over D \phi_i} = 0 \quad {\rm and}
\quad {D \tilde W \over D \sigma_i} = 0,}
then Type IIA compactification on the corresponding Calabi-Yau manifold
is supersymmetric \refs{\GVW, \Gukov}. {}From the formulas \euler,
\sethrel\ and the quantization condition of the $G$-flux \Wflux\ it
follows that there is a finite number of choices for $[G] \in H^4(X)$
corresponding to supersymmetric vacua. In particular, if $h^{2,1} > 8 +
h^{1,1} + h^{3,1}$, then there are no such vacua at all.

{}From the superspace construction in section 5 it follows
that in the equations \minimum\ we should use
the appropriate covariant derivatives:
\eqn\covd{ {D W \over D \phi_i} = {\p W \over \p \phi_i}
+ {\p K_c(\phi_{i}, \bar \phi_{\bar i}) \over \p \phi_i} W,
\quad {D \tilde W \over D \sigma_i} =
{\p \tilde W \over \p \sigma_i} +
{\p K_{\CK} (\s_{j}, \bar \s_{j}) \over \p \sigma_i} \tilde W}
where $K_c$ and $K_{\CK}$ are given by the tree-level formulas
\kphi\ and \ksig, respectively. A simple way to see that one has
to use the covariant derivatives instead of ordinary ones is to
consider first compactification of F-theory on the same Calabi-Yau
space\foot{Of course, here we assume that $X$ is elliptically
fibered. The result, however, is independent of this assumption.}
$X$. In the component action of the effective $\CN=1$ four-dimensional
theory there is a scalar potential:
\eqn\scal{e^K \Big( \CG^{\phi_i \bar \phi_{\bar j}}
(D_{\phi_i} W) (D_{\bar \phi_{\bar j}} \bar W)
- 3 \vert W \vert^2 \Big) }
where the covariant derivative $D_{\phi_i} W = {DW \over D \phi_i}$
is defined in \covd. After a further compactification on a torus
$T^2$, this theory is dual to compactification of Type IIA string
theory on $X$. It is clear that after the dimensional reduction of
the four-dimensional component action the covariant derivatives \covd\
also appear in the component action of the two-dimensional theory in
question.

These models can have a variety of $T$-duality symmetries.
Of particular interest are mirror symmetries \refs{\LVW,\Dixon}.
A mirror symmetry is, of course, a symmetry that
maps Type IIA string theory on a four-fold $X$ to Type
IIA string theory on the mirror variety $\tilde X$, such that:
\eqn\hmirror{ h^{p,q} (X) = h^{4-p,q} (\tilde X)}
and the conformal field theories associated with
$X$ and $\tilde X$ are equivalent.
This operation corresponds \Gates\ to a transformation
which exchanges chiral multiplets and twisted chiral multiplets.
It can be interpreted \GGW\ in terms of
the supergeometrical coordinate transformation
$\theta^{-} \leftrightarrow \theta^{\dot -}$ that also exchanges
chiral multiplets and twisted chiral multiplets and changes the
superspace measure in a way consistent with other definitions of
mirror symmetry. In particular, the latter implies that under the
mirror symmetry we have $\s_i \leftrightarrow \phi_j$ which is
consistent with our interpretation of (vevs of) these fields as the
K\"ahler and the complex structure moduli of the Calabi-Yau space
$X$.  Therefore, the mirror symmetry relates different quantum
$\CN=(2,2)$ theories also interchanging:
$$
U(1)_A \leftrightarrow U(1)_V
$$
\eqn\mirror{ \phi_i \leftrightarrow \s_j }
$$
W(\phi_i) \leftrightarrow \tilde W (\s_j)
$$
The mirror map has no effect on the $\CN=(2,2)$ dilaton
supergravity itself, so that in the absence of matter fields
it must be mirror-symmetric.
It may seem that twisted chiral fields $z_k$ violate
the invariance under \mirror.
Recall that via a spacetime T-duality transformation, those fields
can be described by either chiral or twisted chiral superfields.
Hence mirror symmetry just exchanges these two descriptions.

By definition, the low-energy effective action \action\ describes
dynamics of the light modes in Type IIA string theory on $X$ in the
large volume limit.  In other words, tree-level amplitudes in Type
IIA string theory must agree with the corresponding amplitudes in
the effective two-dimensional theory. In the appendix A we illustrate
this by a world-sheet calculation which independently proves that the
target space metric is block-diagonal, {\it cf.} \split.

A superspace formulation of $\CN=(2,2)$ dilaton supergravity that
includes chiral and twisted chiral multiplets on equal footing does
not seem to exist in the literature. Although a superspace model of
$\CN=(2,2)$ supergravity where the dilaton is a complex field was
constructed in \GKNO, we are interested in a theory where
the supergravity multiplet contains a real dilaton field.
A superspace formulation of such a supergravity
theory is presented in the next section.


\newsec{Superspace Formulation of $\CN=(2,2)$ Dilaton Supergravity}

In this section we present a superspace construction of $\CN=(2,2)$
dilaton supergravity without gauged symmetry. This last property
is a distinguishing feature of the new formulation since all the
known $\CN=(2,2)$ gravity theories have at least one gauged $U(1)$
$R$-symmetry (see \GGW\ for a general presentation).
Theories where the entire $U(1)_A \otimes U(1)_V$ symmetry
group is gauged are called non-minimal (or reducible), as opposed
to minimal theories where only $U(1)_A$ or $U(1)_V$ factor is gauged.
It is very well known how to obtain one supergravity theory
with a smaller holonomy group from a supergravity theory
with a larger holonomy group. This process has been for
a long time called ``de-gauging'' (see \GGRS, section 5.3.b.7).

The basic idea of de-gauging is to break the gauge symmetry
introducing extra matter field in a Goldstone-like mechanism.
For example, consider an abelian vector multiplet
in a four-dimensional $\CN=1$ gauge theory.
It contains a $U(1)$ gauge vector field, gaugino and an auxiliary field.
All the other fields can be set to zero by a supersymmetric
choice of gauge, the so-called Wess-Zumino gauge.
On the other hand, a massive gauge multiplet contains
some extra component fields which could be eliminated
in the massless multiplet. The reason is that the massive
system no longer possess the $U(1)$ gauge invariance.
This toy model teaches us that when a symmetry is broken
in superspace, extra component fields not present in the gauge
symmetric phase begin to appear. In other words, Goldstone
supermultiplets must appear. And their component
fields come from that part of the vector multiplet
that was ignored in the symmetric phase.
Following these steps,
we construct $\CN=(2,2)$ dilaton supergravity via de-gauging
$U(1)_A \otimes U(1)_V$ non-minimal gauged supergravity.
An advantage of this approach is that both the original and
the resulting theories are manifestly invariant under
the mirror symmetry \mirror. We also find that
the new $\CN=(2,2)$ supergravity multiplet contains a real dilaton
field $\varphi$, in accordance with the results  of section 3 where
we studied compactification of Type IIA string theory on Calabi-Yau
four-folds. We hope that apart from this obvious application there
may also be many other aspects of the new supergravity to explore.
For example, it would be interesting to study black hole
solutions in this dilaton supergravity, {\it cf.} \CGHS.

The relation between different supergravity theories can be
schematically represented in the form of the following diagram:
\eqn\sugras{ \matrix{
& & U(1)_A \otimes U(1)_V & & \cr
& \swarrow & & \searrow & \cr
U(1)_A & & \downarrow & & U(1)_V \cr
& \searrow & & \swarrow & \cr
& & {\bf 1} & &}}
where the theory with the trivial holonomy group
is the new $\CN=(2,2)$ dilaton supergravity we are going
to construct. Notice, however, that various arrows in this
diagram have different meaning. For example, minimal theories
with either of the $U(1)$ R-symmeties gauged can be
obtained by truncation of the non-minimal $U(1)_A \otimes U(1)_V$
gauged supergravity \GGW. On the other hand, the vertical arrow
corresponds to de-gauging $U(1)_A \otimes U(1)_V$ symmetry,
so that the total number of degrees of freedom increases.
More precisely, it has to be a combination of consistent
truncation of the non-minimal $\CN=(2,2)$ supergravity
to a minimal one plus a de-gauging of the latter.
To see this, let us count the number of real Goldstone scalars.
Since the broken $U(1)_A \otimes U(1)_V$ phase of the theory
has exactly the same field content as the gauge symmetric phase
plus the field content of the Goldstone multiplets minus
the parts that go into the longitudinal components of
the $U(1)_A \otimes U(1)_V$ gauge fields, we find that
in total Goldstone multiplets should have three real scalars.
However, there are no mirror-symmetric $\CN=(2,2)$ multiplets
with such a field content. Therefore, we conclude that
the vertical arrow should be a more economical de-gauging.
Indeed, if we were following another route via a minimal
gauged supergravity, at the first step we would have
to make a consistent truncation that would eliminate one
of the gauge fields. In order to de-gauge the resulting minimal
gauged supergravity we would have to introduce extra
chiral (or twisted chiral) Goldstone superfield.
In any case, one real scalar from this multiplet
would become the longitudinal
component of the gauge vector field, and the other would
become a dilaton, in agreement with what we expect.
Assuming that the latter route (which is not, unfortunately,
mirror-symmetric) is equivalent to the vertical arrow on
the above diagram, we expect that there is an economical
de-gauging of the non-minimal $\CN=(2,2)$ supergravity
that leads to only one massless scalar (the dilaton).

To start the construction, let us arrange component fields found
in the  previous section into superfields.
First we give the definitions in flat superspace
and then extend them to curved superspace.
Left-right symmetric  $\CN=2$ superspace is parametrized by
the bosonic coordinates $x^{\mu} = (x^{\pp},  x^{\mm})$, two
anti-commuting complex spinor coordinates
$\th^{\a} = (\th^{+}, \th^{-})$ and their complex conjugates $\bar
\th^{\dot \a} = (\bar \th^{\dot +}, \bar \th^{\dot -})$. The spinor
derivatives $D_{+}$, $D_{-}$, $D_{\dot +}$ and $D_{\dot -}$ satisfy
$\{ D_{+}, D_{\dot +} \} = \p_{\pp}$ and $\{ D_{-}, D_{\dot -} \} =
\p_{\mm}$, with all other (anti-)commutators vanishing.

Irreducible matter superfields are defined by imposing some constraints
on general complex superfields. The simplest constraints are linear in
derivatives and look like:
\eqn\cconstr{D_{\dot +} \Phi = D_{\dot -} \Phi =
D_{+} \overline \Phi = D_{-} \overline \Phi = 0}
for a chiral superfield $\Phi$, and:
\eqn\tconstr{D_{\dot +} \S = D_{-} \S =
D_{+} \overline \S = D_{\dot -} \overline \S = 0}
for a twisted chiral superfield $\S$.

In what follows we promote the complex scalar fields $\phi_i$ and $\s_j$
defined in the previous section to the chiral and twisted chiral
superfields $\Phi_i$ and $\S_j$, respectively.
Similarly, we will regard the compact fields $z_k$ as
the scalar components of (twisted) chiral superfields $\CZ_k$.

Components of the chiral superfield $\Phi_i$ can be obtained using the
projection method:
$$
\Phi_i \vert = \phi_i, \quad \overline \Phi_i \vert = \overline \phi_i
$$
\eqn\cmult{D_{+} \Phi_i \vert = \psi^i_{+}, \quad
D_{\dot +} \overline \Phi_i \vert = \psi^i_{\dot +} }
$$
D_{-} \Phi_i \vert = \psi^i_{-}, \quad D_{\dot -}
\overline \Phi_i \vert = \psi^i_{\dot -}
$$
$$
{i \over 2} [D_{+}, D_{-}] \Phi_i \vert = A_i, \quad {i \over 2} [
D_{\dot +}, D_{\dot -}] \overline \Phi_i \vert = \overline A_i
$$
where, for example, $\Phi_i \vert$ denotes the leading component of
the superfield $\Phi_i$, with all the $\th$-coordinates put to zero.
Similarly, we find the components of a twisted chiral multiplet $\S_j$:
$$
\S_j \vert = \s_j, \quad \overline \S_j \vert = \overline \s_j
$$
\eqn\tmult{ D_{+} \S_j \vert = \zeta^j_{+}, \quad
D_{\dot +} \overline \S_j \vert = \zeta^j_{\dot +} }
$$
D_{\dot -} \S_j \vert = \zeta^j_{\dot -}, \quad
D_{-} \overline \S_j \vert = \zeta^j_{-}
$$
$$
{i \over 2} [D_{+}, D_{\dot -}] \S_j \vert = B_j, \quad {i \over
2} [D_{\dot +}, D_{-}] \overline \S_j \vert = \overline B_j
$$
To define chiral and twisted chiral superfields in curved $\CN=(2,2)$
superspace, the spinor derivatives $D_{\a}$ must be appropriately
replaced by the covariant derivatives $\nabla_{\a}$.

In our construction of $\CN=(2,2)$ dilaton supergravity we start
with non-minimal gauged supergravity and then de-gauge
$U(1)_A \otimes U(1)_V$ symmetry.
If we introduce superfields $\CA_{\a}$ and $\CA_{\a}'$
representing $U(1)_V$ and $U(1)_A$ gauge connections,
and denote by $\Lambda_{\a}$ the Lorentz spin-connection,
then the covariant derivative in this theory has the form:
\eqn\cderiv{ \nabla_{\a} = {E_{\a}}^{B} D_{B} + \Lambda_{\a}
\CX + \CA_{\a} \CY + \CA_{\a}' \CY'}
where ${E_{\a}}^{B}$ is the supervielbein.

The Lorentz generators, $\CX$,
$U(1)_V$ symmetry generators, $\CY$,
and $U(1)_A$ symmetry generators, $\CY'$,
act on the covariant derivative $\nabla_{\a}$ in the following way \GGW:
$$
[ \CX, \nabla_{\pm} ] = \pm \half \nabla_{\pm} \quad , \quad
[ \CX, \nabla_{\dot \pm} ] = \pm \half \nabla_{\dot \pm}
$$
\eqn\ycommut{[ \CY, \nabla_{\pm} ] = - {i \over 2} \nabla_{\pm}
\quad , \quad [ \CY, \nabla_{\dot \pm} ] = + {i \over 2}
\nabla_{\dot \pm} }
$$
[ \CY', \nabla_{\pm} ] = \mp {i \over 2} \nabla_{\pm}
\quad , \quad [ \CY', \nabla_{\dot \pm} ] = \pm {i \over 2}
\nabla_{\dot \pm}
$$

The constraints which define non-minimal $\CN=2$
$U(1)_A \otimes U(1)_V$ supergravity are given by:
$$
\{ \nabla_{+}, \nabla_{+}\} = 0, \quad \{ \nabla_{-}, \nabla_{-}\} = 0
$$
\eqn\sugra{ \{ \nabla_{+}, \nabla_{\dot +}\} = i \nabla_{\pp},
\quad \{ \nabla_{-}, \nabla_{\dot -}\} = i \nabla_{\mm} }
$$
\{ \nabla_{+}, \nabla_{-} \} =
- {1 \over 2} \bar R (\bar \CX - i \bar \CY') ,
\quad \{ \nabla_{+}, \nabla_{\dot -}\} =
- {1 \over 2} \bar F (\bar \CX - i \bar \CY)
$$
where the chiral superfield $R$ and the twisted chiral superfield $F$
are related to the two-dimensional curvature $R^{(2)}$ and the abelian
field strengths of the graviphoton gauge fields.
The easiest way to see this is to compute the commutator \refs{\GGW,\A}:
\eqn\curvcomm{
[\nabla_{\pp}, \nabla_{\mm}] =
{1 \over 2} \Big( (\nabla^2 R) - {1 \over 2} R \overline R +
(\nabla_{+} \nabla_{\dot -} F) - {1 \over 2} F \overline F \Big) \CX +}
$$
+ {i \over 2} (\nabla_{+} \nabla_{\dot -} F) \CY +
{i \over 2} (\nabla^2 R) \CY'+ \ldots + {\rm c.c.}
$$
where the dots stand for the covariant derivative terms like
$(\nabla_{-} F) \nabla_{\dot +}$, {\it etc.}
If we set $F=0$ in the constraints \sugra, we obtain $U(1)_A$ minimal
gauged supergravity theory.
On the other hand, if we set $R=0$, we end up
with $U(1)_V$ minimal theory.
We denote the leading components of the superfields
$F$ and $R$ as follows:
\eqn\fgrh{ F \vert = G, \quad R \vert = H }

In order to obtain $\CN=(2,2)$ supergravity theory without gauged
symmetry, we replace the covariant derivative $\nabla_{\a}$ by
$\hat \nabla_{\a}$ which includes only derivatives and Lorentz
generator, so that:
\eqn\newderiv{\nabla_{\a} = \hat \nabla_{\a} + \CA_{\a} \CY
+ \CA_{\a}' \CY'}
and the remaining derivatives are real.
The new covariant derivative $\hat \nabla_{\a}$ contains
neither the $U(1)_V$ gauge connection nor the $U(1)_A$
gauge connection, so it can describe two-dimensional
$\CN=(2,2)$ supergravity without gauged symmetry.

Substituting \newderiv\ into \sugra, and using \ycommut, we derive the
form of the commutator algebra for the $\hat \nabla_{\a}$ operators:
$$
\{ \hat \nabla_{+}, \hat \nabla_{+}\} =
i (\la_{+} + \tilde \la_{+}) \hat \nabla_{+}
\quad , \quad
\{ \hat \nabla_{-}, \hat \nabla_{-}\} =
i (\la_{-} - \tilde \la_{-}) \hat \nabla_{-}
$$
$$
\{ \hat \nabla_{+}, \hat \nabla_{-} \} = - {1 \over 2} \bar R \CX
+ {i \over 2} (\la_{-} + \tilde \la_{-}) \hat \nabla_{+}
+ {i \over 2} (\la_{+} - \tilde \la_{+}) \hat \nabla_{-}
$$
\eqn\newsugra{
\{ \hat \nabla_{+}, \hat \nabla_{\dot -} \} = - {1 \over 2} \bar F \CX
+ {i \over 2} (\la_{\dot -} + \tilde \la_{\dot -}) \hat \nabla_{+}
- {i \over 2} (\la_{+} - \tilde \la_{+}) \hat \nabla_{\dot -}}
$$
\{ \hat \nabla_{+}, \hat \nabla_{\dot +}\} = i \hat \nabla_{\pp}
+ {i \over 2} (\la_{\dot +} + \tilde \la_{\dot +}) \hat \nabla_{+}
- {i \over 2} (\la_{+} + \tilde \la_{+}) \hat \nabla_{\dot +}
$$
$$
\{ \hat \nabla_{-}, \hat \nabla_{\dot -}\} = i \hat \nabla_{\mm}
+ {i \over 2} (\la_{\dot -} - \tilde \la_{\dot -}) \hat \nabla_{-}
- {i \over 2} (\la_{-} - \tilde \la_{-}) \hat \nabla_{\dot -}
$$
where the new fields $\la_{\a}$ and $\tilde \la_{\a}$ have appeared.
They are components of the non-minimal
gauged supergravity multiplet that could be eliminated by a gauge
transformation in the $U(1)_A \otimes U(1)_V$ gauge-symmetric phase.
In the theory we are constructing the $U(1)_A \otimes U(1)_V$
gauge symmetry is broken, so that the fields $\la_{\a}$
and $\tilde \la_{\a}$ are dynamical.
They have to be identified with the spinorial derivatives
of a new matter Goldstone multiplet.
To the leading order in $\theta^{\a}$, the fields
$\la_{\a}$ and $\tilde \la_{\a}$ can also be identified
with the dilatino field of the new supergravity multiplet.

If we introduce four linear independent spinors
(along with their conjugates):
$$
\eta_{+} \equiv \la_{+} + \tilde \la_{+}, \quad
\eta_{-} \equiv \la_{-} - \tilde \la_{-},
$$
$$
\tilde \eta_{+} \equiv \la_{+} - \tilde \la_{+}, \quad
\tilde \eta_{-} \equiv \la_{-} + \tilde \la_{-}
$$
then the conditions \newsugra\ which define $\CN=(2,2)$
dilaton supergravity can be written in the following simple form:
$$
\{ \hat \nabla_{+}, \hat \nabla_{+}\} = i \eta_{+} \hat \nabla_{+}
\quad , \quad \{ \hat \nabla_{-}, \hat \nabla_{-}\} =
i \eta_{-} \hat \nabla_{-}
$$
$$
\{ \hat \nabla_{+}, \hat \nabla_{-} \} = - {1 \over 2} \bar R \CX
+ {i \over 2} \tilde \eta_{-} \hat \nabla_{+}
+ {i \over 2} \tilde \eta_{+} \hat \nabla_{-}
$$
$$
\{ \hat \nabla_{+}, \hat \nabla_{\dot -} \} = - {1 \over 2} \bar F \CX
+ {i \over 2} \tilde \eta_{\dot -} \hat \nabla_{+}
- {i \over 2} \tilde \eta_{+} \hat \nabla_{\dot -}
$$
$$
\{ \hat \nabla_{+}, \hat \nabla_{\dot +}\} = i \hat \nabla_{\pp}
+ {i \over 2} \eta_{\dot +} \hat \nabla_{+}
- {i \over 2} \eta_{+} \hat \nabla_{\dot +}
$$
$$
\{ \hat \nabla_{-}, \hat \nabla_{\dot -}\} = i \hat \nabla_{\mm}
+ {i \over 2} \eta_{\dot -} \hat \nabla_{-}
- {i \over 2} \eta_{-} \hat \nabla_{\dot -}
$$
Note that these equations, as well as the conditions \newsugra,
are manifestly invariant under the mirror symmetry transformation \mirror.

To summarize, there exists a unique mirror-symmetric two-dimensional
$\CN = (2,2)$ dilaton supergravity defined by the set of constraints
\newsugra\ on the covariant derivative $\hat \nabla_{\a}$.
The new supergravity theory does not have gauged symmetry and
contains a real dilaton field $\varphi$, as we will show in a moment.

Since we define new $\CN=(2,2)$ dilaton supergravity theory imposing
constraints \newsugra\ on the covariant derivatives $\hat \nabla_{\a}$,
the Bianchi identities in this theory may lead to further constraints
on some fields\foot{In ordinary field theories, the fields satisfy
Bianchi identities because they are expressed in terms of the potentials;
they are {\it identities} and impose no extra constraints.}.
For theories described by covariant derivatives $\hat \nabla_{\a}$,
the Bianchi identities are simply Jacobi identities:
\eqn\jacobi{ [ \hat \nabla_{ [ \a}, [\hat \nabla_{\b},
\hat \nabla_{ \gamma )} \} \} = 0 }
where $[~ ,~ \}$ is the graded commutator, and $[~ ,~ )$ stands
for the graded antisymmetrization symbol.
Instead of deriving derivative constraints on the spinor fields
$\la_{\a}$ and $\tilde \la_{\a}$ directly from the Jacobi identities
\jacobi\ we use an equivalent approach which is much easier.
While substituting \newderiv\ into \sugra\ one also finds terms
proportional to gauge symmetry generators $\CY$ and $\CY'$.
Vanishing of these terms leads to a set of constraints which
is equivalent to the set of constraints obtained from
the Jacobi identities \jacobi.
We outline the result in appendix B.

There are two simple solutions to the Jacobi identities corresponding
to either $\eta_{\a}$ or $\tilde \eta_{\a}$ put to zero.
An advantage of the first solution is that the covariant derivative
$\hat \nabla_{\a}$ anti-commutes with itself, $\hat \nabla_{\a}^2 =0$,
like in the usual gauged supergravity theories \GGW.
On the other hand, in the second case we find especially simple form
of the anti-commutators $\{ \hat \nabla_{+}, \hat \nabla_{-} \}$
and $\{ \hat \nabla_{+}, \hat \nabla_{\dot -} \}$.
In both cases the remaining spinor superfields can be
expressed in terms of an unconstraint real superfield $V$:
\eqn\vreal{V = \bar V}
so that the Jacobi identities \jacobi\ are satisfied.
This means that \jacobi\ impose no further constraints
on $V$, and only define the other superfields
(like $F$ and $R$) in terms of the derivatives of $V$.
It is natural to identify the dilaton field with
the leading scalar component of $V$:
\eqn\vdilaton{\varphi = V \vert}
Below we present more evidence for this identification.
One might notice that a real superfield $V$ contains
one massless vector field, in agreement with the result of
the previous section\foot{However, massless vector fields in
two dimensions do not have propagating degrees of freedom.
For the same reason two-dimensional superfield $V$ does not
have an irreducible transverse component, unlike a similar
four-dimensional superfield.}.
It is also worthwhile to stress here that massless
superfield $V$ is not a Goldstone multiplet itself,
but rather what remains after the Goldstone mechanism
takes place. A nice property of this solution is that
$V$ is manifestly mirror-symmetric.

Since local integration measures of the new $\CN=(2,2)$
dilaton supergravity can be nicely derived from the
corresponding expressions of the $U(1)_A \otimes U(1)_V$
theory only for the solution corresponding to $\tilde \eta_{\a}=0$,
in what follows we discuss in detail only this case.
Namely, we take the following ansatz for the spinors $\la_{\a}$:
\eqn\vla{\la_{+} = \tilde \la_{+} = i (\hat \nabla_{+} V)
\quad , \quad \la_{-} = - \tilde \la_{-} = i (\hat \nabla_{-} V)}
$$
\la_{\dot +} = \tilde \la_{\dot +} = - i (\hat \nabla_{\dot +} V)
\quad , \quad
\la_{\dot -} = - \tilde \la_{\dot -} = - i (\hat \nabla_{\dot -} V)
$$
which implies $\eta_{\a} = 2i \hat \nabla_{\a} V$
and $\tilde \eta_{\a} = 0$.

Substituting \vla\ into \newsugra, we find the following
supergravity algebra:
$$
\{ \hat \nabla_{+}, \hat \nabla_{+}\} =
- 2 (\hat \nabla_{+} V) \hat \nabla_{+}
\quad , \quad
\{ \hat \nabla_{-}, \hat \nabla_{-}\} =
- 2 (\hat \nabla_{-} V) \hat \nabla_{-}
$$
\eqn\vsugra{ \{ \hat \nabla_{+}, \hat \nabla_{-} \} =
- {1 \over 2} \bar R \CX \quad , \quad
\{ \hat \nabla_{+}, \hat \nabla_{\dot -} \} = - {1 \over 2} \bar F \CX }
$$
\{ \hat \nabla_{+}, \hat \nabla_{\dot +}\} = i \hat \nabla_{\pp}
+ (\hat \nabla_{\dot +} V) \hat \nabla_{+}
+ (\hat \nabla_{+} V) \hat \nabla_{\dot +}
$$
$$
\{ \hat \nabla_{-}, \hat \nabla_{\dot -}\} = i \hat \nabla_{\mm}
+ (\hat \nabla_{\dot -} V) \hat \nabla_{-}
+ (\hat \nabla_{-} V) \hat \nabla_{\dot -}
$$
where the superfields $\bar R$ and $\bar F$ can
be obtained from the Bianchi identities.
Solving the set of constraints in appendix B we get:
\eqn\vrf{ \bar R = 4 \hat \nabla_{-} \hat \nabla_{+} V
\quad , \quad \bar F = 4 \hat \nabla_{\dot -} \hat \nabla_{+} V}
We also find the following expressions for the gauge connection:
$$
\la_{\pp} = \tilde \la_{\pp} =
i \hat \nabla_{\pp} V - 2 \hat \nabla_{+} \hat \nabla_{\dot +} V
+ 4 (\hat \nabla_{+} V) (\hat \nabla_{\dot +} V)
$$
$$
\la_{\mm} = - \tilde \la_{\mm} =
i \hat \nabla_{\mm} V - 2 \hat \nabla_{-} \hat \nabla_{\dot -} V
+ 4 (\hat \nabla_{-} V) (\hat \nabla_{\dot -} V)
$$

Further commutators of the covariant derivatives
with vector indices follow from the consistency
of the Bianchi identities \jacobi:
$$
[\hat \nabla_{+}, \hat \nabla_{\pp}] =
- \Big( 2i \la_{+} \la_{\dot +} + (\hat \nabla_{\dot +} \la_{+})
+ (\hat \nabla_{+} \la_{\dot +})\Big) \hat \nabla_{+}
$$
$$
[\hat \nabla_{\dot +}, \hat \nabla_{\pp}] =
\Big( 2i \la_{+} \la_{\dot +} + (\hat \nabla_{\dot +} \la_{+})
+ (\hat \nabla_{+} \la_{\dot +})\Big) \hat \nabla_{\dot +}
$$
$$
[\hat \nabla_{-}, \hat \nabla_{\mm}] =
- \Big( 2i \la_{-} \la_{\dot -} + (\hat \nabla_{\dot -} \la_{-})
+ (\hat \nabla_{-} \la_{\dot -})\Big) \hat \nabla_{-}
$$
\eqn\newcommut{ [\hat \nabla_{\dot -}, \hat \nabla_{\mm}] =
\Big( 2i \la_{-} \la_{\dot -} + (\hat \nabla_{\dot -} \la_{-})
+ (\hat \nabla_{-} \la_{\dot -})\Big) \hat \nabla_{\dot -} }
$$
 [\hat \nabla_{+}, \hat \nabla_{\mm}] =
- \Big( {i \over 2} (\hat \nabla_{\dot -} \bar R)
+ {i \over 2} (\hat \nabla_{-} \bar F)
+ \half \la_{\dot -} \bar R - \half \la_{-} \bar F \Big) \CX
- {i \over 2} \bar R \hat \nabla_{\dot -}
- {i \over 2} \bar F \hat \nabla_{-}
$$
$$
[\hat \nabla_{\dot +}, \hat \nabla_{\mm}] =
\Big( {i \over 2} (\hat \nabla_{-} R)
+ {i \over 2} (\hat \nabla_{\dot -} F)
- \half \la_{-} R + \half \la_{\dot -} F \Big) \CX
+ {i \over 2} R \hat \nabla_{-}
+ {i \over 2} F \hat \nabla_{\dot -}
$$
$$
[\hat \nabla_{-}, \hat \nabla_{\pp}] =
- \Big( {i \over 2} (\hat \nabla_{\dot +} \bar R)
+ {i \over 2} (\hat \nabla_{+} F)
+ \half \la_{\dot +} \bar R - \half \la_{+} F \Big) \CX
+ {i \over 2} \bar R \hat \nabla_{\dot +}
+ {i \over 2} F \hat \nabla_{+}
$$
$$
[\hat \nabla_{\dot -}, \hat \nabla_{\pp}] =
\Big( {i \over 2} (\hat \nabla_{+} R)
+ {i \over 2} (\hat \nabla_{\dot +} \bar F)
- \half \la_{+} R + \half \la_{\dot +} \bar F \Big) \CX
- {i \over 2} R \hat \nabla_{+}
- {i \over 2} \bar F \hat \nabla_{\dot +}
$$
where we used \vsugra.
It is worthwhile to stress here that one would obtain
a different result de-gauging the corresponding commutators
in the $U(1)_A \otimes U(1)_V$ non-minimal supergravity \GW.


\newsec{Lagrangians for Matter Multiplets Coupled to
$\CN=2$ Dilaton Supergravity}

In order to couple matter fields to new $\CN=(2,2)$ dilaton
supergravity we have to repeat the analysis of \GGW.
Up to terms with two derivatives or four fermions, the most general
action of $\CN=(2,2)$ supergravity coupled to chiral superfields
$\Phi_i$ and $\CZ_k$, and twisted chiral superfields $\S_j$
looks like \action:
\eqn\superaction{ S = \int d^2x \int  d^2 \th d^2 \bar \th
E^{-1} \CL (\Phi_{i}, \bar \Phi_{\bar i}, \S_{j}, \bar \S_{j},
\CZ_k, \bar \CZ_{\bar k} ) +}
$$
+ \int d^2x \int d^2 \th \CE^{-1} W(\Phi_i) +
\int d^2x \int d \th^{+} d \th^{\dot -} \tilde \CE^{-1}
\tilde W(\S_j) + {\rm c.c.}
$$

In order to obtain the component action corresponding to \superaction,
one needs the appropriate projection formulas.
For gauged $\CN=(2,2)$ supergravity theories such formulas
were derived by Grisaru and Wehlau \GWmeasure.
In the case of the minimal $U(1)_A$ theory the local density
projection formula has the following form:
\eqn\cproj{\int d^2x d^4 \th E^{-1} \CL =
\int d^2x e^{-1} \big[ \nabla^2
+ i \psi^{\dot -}_{\mm} \nabla_{+}
- i \psi^{\dot +}_{\pp} \nabla_{-} + }
$$
+ (- {1 \over 2} \overline H
- \psi^{\dot -}_{\pp} \psi^{\dot +}_{\mm}
+ \psi^{\dot -}_{\mm} \psi^{\dot +}_{\pp} )\big]
\overline \nabla^2 \CL \vert
$$
Here $\psi^{\a}_{\mu}$ is the gravitino field and $\CL$ is an
arbitrary scalar function of superfields.
In fact, the same projection formula is also valid in
the non-minimal $U (1)_A \otimes U (1)_V$ supergravity theory \GGW.
Although \cproj\ is a D-type superinvariant, sometimes
it is called a chiral density projector because in the non-minimal
$\CN=(2,2)$ supergravity $\overline \nabla^2 \CL$
is a chiral superfield (for a general $\CL$).
Therefore, replacing $\overline \nabla^2 \CL$ by an arbitrary
covariantly chiral Lagrangian $\CL_c$, we can obtain the component
projection formula for any chiral superspace integral:
\eqn\lcproj{\int d^2x d^2 \th \CE^{-1} \CL_c =
\int d^2x e^{-1} \big[ \nabla^2
+ i \psi^{\dot -}_{\mm} \nabla_{+}
- i \psi^{\dot +}_{\pp} \nabla_{-} + }
$$
+ (- {1 \over 2} \overline H
- \psi^{\dot -}_{\pp} \psi^{\dot +}_{\mm}
+ \psi^{\dot -}_{\mm} \psi^{\dot +}_{\pp} )\big] \CL_c \vert
$$
In particular, the superspace measures $E^{-1}$ and $\CE^{-1}$
are related as follows:
\eqn\measure{\int d^2x d^4 \th E^{-1} \CL =
\int d^2x d^2 \th \CE^{-1} \overline \nabla^2 \CL \vert}

By mirror symmetry the twisted chiral density projection formula in
the $U(1)_V$ gauged supergravity theory has the following form:
\eqn\tproj{ \int d^2x d^4 \th E^{-1} \CL =
\int d^2x e^{-1} \big[ \nabla_{\dot -} \nabla_{+} -
i \psi^{-}_{\mm} \nabla_{+} + i \psi^{\dot +}_{\pp} \nabla_{\dot -} +}
$$
+ (- {1 \over 2} \overline G + \psi^{-}_{\pp} \psi^{\dot +}_{\mm} +
\psi^{\dot +}_{\pp} \psi^{-}_{\mm} )\big]
\nabla_{\dot +} \nabla_{-} \CL \vert
$$
As explained in \GGW, the derivation of this formula
goes through as in \GWmeasure\ for the case of $U (1)_A$ theory.
In the case of the $U (1)_V \otimes U (1)_A$ gauged supergravity
the symmetry between chiral and twisted chiral fields
is restored by the contribution of the anticommutator term
$\{ \nabla_{+}, \nabla_{\dot -} \} \sim \overline F \vert$.
More explicitly, the twisted chiral density projection
formula \tproj\ can be derived using the methods
of \GWmeasure\ or \G.

The projection formulas in the two-dimensional
$\CN=(2,2)$ dilaton supergravity can be obtained from
\cproj\ and \tproj\ replacing $\nabla_{\a}$ by a new covariant
derivative $\hat \nabla_{\a}$.
Thus, substituting \newderiv\ and \vla\ into \cproj\ and using
the commutation relations \ycommut\ -- \sugra\ we obtain
the following density projection formula:
$$
\int d^2x d^4 \th E^{-1} \CL = \int d^2x e^{-1} \big[
(\hat \nabla_{+} - (\hat \nabla_{+} V))
(\hat \nabla_{-} - (\hat \nabla_{-} V))
+ i \psi^{\dot -}_{\mm} (\hat \nabla_{+} - (\hat \nabla_{+} V)) -
$$
\eqn\newcproj{
- i \psi^{\dot +}_{\pp} (\hat \nabla_{-} - (\hat \nabla_{-} V))
+ (- {1 \over 2} \overline H
- \psi^{\dot -}_{\pp} \psi^{\dot +}_{\mm}
+ \psi^{\dot -}_{\mm} \psi^{\dot +}_{\pp} )\big]
\hat \nb^2 \CL \vert = }
$$
= \int d^2x e^{-1} \big[ \hat \nabla_{+} \hat \nabla_{-}
+ i (\psi_{\mm}^{\dot -} - \la_{-}) \hat \nabla_{+}
- i (\psi_{\pp}^{\dot +} - \la_{+}) \hat \nabla_{-} +
$$
$$
+ \big( - {1 \over 4} \bar H
- \psi_{\pp}^{\dot -} \psi_{\mm}^{\dot +}
+ (\psi_{\mm}^{\dot -} - \la_{-})
(\psi_{\pp}^{\dot +} - \la_{+})
\big) \big] \hat \nabla_{\dot +} \hat \nabla_{\dot -} \CL \vert
$$
In order to convince even hard boiled sceptics that \newcproj\
is the right projector, in appendix D we repeat the calculation
of Grisaru and Wehlau \GWmeasure\ in the new $\CN=(2,2)$ dilaton
supergravity. As expected, the result is equivalent to \newcproj.

Similarly, the twisted chiral density projection formula \tproj\ yields:
$$
\int d^2x d^4 \th E^{-1} \CL =
\int d^2x e^{-1} \big[
(\hat \nabla_{\dot -} - (\hat \nabla_{\dot -} V))
(\hat \nabla_{+} - (\hat \nabla_{+} V))
- i \psi^{-}_{\mm}
(\hat \nabla_{+} - (\hat \nabla_{+} V)) +
$$
\eqn\newtproj{ + i \psi^{\dot +}_{\pp}
(\hat \nabla_{\dot -} - (\hat \nabla_{\dot -} V))
+ (- {1 \over 2} \overline G + \psi^{-}_{\pp} \psi^{\dot +}_{\mm} +
\psi^{\dot +}_{\pp} \psi^{-}_{\mm} )\big]
\hat \nabla_{\dot +} \hat \nabla_{-} \CL \vert }
We note that for a given superspace Lagrangian $\CL$,
both projection formulas \newcproj\ and \newtproj\
lead to the same result:
$$
\int d^2x d^4 \th E^{-1} \CL = \int d^2x e^{-1} \big[
(\hat \nabla_{+} - (\hat \nabla_{+} V))
(\hat \nabla_{-} - (\hat \nabla_{-} V))
+ i \psi^{\dot -}_{\mm} (\hat \nabla_{+} - (\hat \nabla_{+} V)) -
$$
$$
- i \psi^{\dot +}_{\pp} (\hat \nabla_{-} - (\hat \nabla_{-} V))
+ (- {1 \over 2} \overline H
- \psi^{\dot -}_{\pp} \psi^{\dot +}_{\mm}
+ \psi^{\dot -}_{\mm} \psi^{\dot +}_{\pp} )\big]
\hat \nb^2 \CL \vert =
$$
$$
= \int d^2x e^{-1} \big[
(\hat \nabla_{\dot -} - (\hat \nabla_{\dot -} V))
(\hat \nabla_{+} - (\hat \nabla_{+} V))
- i \psi^{-}_{\mm}
(\hat \nabla_{+} - (\hat \nabla_{+} V)) +
$$
$$
+ i \psi^{\dot +}_{\pp} (\hat \nabla_{\dot -} - (\hat \nabla_{\dot -} V))
+ (- {1 \over 2} \overline G + \psi^{-}_{\pp} \psi^{\dot +}_{\mm} +
\psi^{\dot +}_{\pp} \psi^{-}_{\mm} )\big]
\hat \nabla_{\dot +} \hat \nabla_{-} \CL \vert
$$
This follows from the corresponding property of the local density
projectors in gauged $\CN=(2,2)$ supergravity theory \GWmeasure,
and also can be verified explicitly using the commutation relations
\vsugra\ in the new $\CN=(2,2)$ dilaton supergravity.

Now we are ready to derive component actions for various
superspace Lagrangians $\CL$. Let us start with a simple
example corresponding to pure dilaton supergravity.
Obviously, in order to reproduce the right exponential
dependence on the dilaton field in \act, we have
to take the function $\CL$ in the form:
$$
\CL_{\rm grav} = \exp (- 2V).
$$
Substituting this in the projection formula \newcproj\
(or \newtproj) and seting all the fermions to zero,
we obtain the following action for bosonic fields:
\eqn\gravact{ S_{\rm grav}
= \int d^2x e^{-1} \exp (-2 \varphi)
\big[ R^{(2)} + 4 (\p_{\mu} \varphi) (\p^{\mu} \varphi) \big] }
Here we used \vla\ and the formulas for the other
components of the real superfield $V$ derived in appendix C.
Clearly, the action \gravact\ for dilaton and graviton fields
agrees\foot{We will account for the extra volume factor $\CV$
in a moment.} with the first two terms in the effective
action \act\ of Type IIA theory on a Calabi-Yau four-fold.
Moreover, we note that bosonic action \gravact\ has exactly
the same form as the action of $\CN=0$ dilaton gravity
studied long time ago, see {\it e.g.} \CGHS.

Now we consider coupling of $\CN=(2,2)$ dilaton supergravity
to matter fields. In particular, we are interested in
superspace form of the effective action \act\ describing
compactification of Type IIA string theory on a Calabi-Yau four-fold $X$.
Once again, to reproduce the exponential dependence on
$\varphi = V \vert$ we take the superspace action in the form:
\eqn\generalact{ \int d^2x d^4 \th E^{-1} \exp (- 2V) \CL}
where $\CL$ is a function of all matter superfields but $V$.
It is convenient to absorb $\exp (- 2V)$ in the definition
of the supervielbein determinant $E_0^{-1} = E^{-1} \exp (- 2V)$,
so that $\CL=1$ corresponds to pure supergravity action \gravact,
like in $\CN=1$ four-dimensional theory.
Commuting $\exp (- 2V)$ to the left in \newcproj,
we find the modified projection formula:
\eqn\newproj{ \int d^2x d^4 \th E_0^{-1} \CL = }
$$
= \int d^2x e^{-1} \exp (-2 \varphi) \big[
(\hat \nabla_{+} - 3 (\hat \nabla_{+} V))
(\hat \nabla_{-} - 3 (\hat \nabla_{-} V))
+ i \psi^{\dot -}_{\mm} (\hat \nabla_{+} - 3 (\hat \nabla_{+} V)) -
$$
$$
- i \psi^{\dot +}_{\pp} (\hat \nabla_{-} - 3 (\hat \nabla_{-} V))
+ (- {1 \over 2} \overline H
- \psi^{\dot -}_{\pp} \psi^{\dot +}_{\mm}
+ \psi^{\dot -}_{\mm} \psi^{\dot +}_{\pp} )\big]
(\hat \nabla_{\dot +} - 2 (\hat \nabla_{\dot +} V))
(\hat \nabla_{\dot -} - 2 (\hat \nabla_{\dot -} V)) \CL \vert
$$
Applying this projection formula to an arbitrary function
$\CL (\Phi_i, \bar \Phi_{\bar i}, \S_j, \bar \S_j)$
of chiral superfields $\Phi_i$ and twisted chiral
superfields $\S_j$ we get the action of the bosonic fields
(the vielbein determinant $e^{-1}$ is suppressed):
$$
L = e^{-2 \varphi} \CL \big[
R^{(2)} + 4 \p_{\mu} (\varphi - \half \log \CL)
\p^{\mu} (\varphi - \half \log \CL) +
$$
\eqn\actcomp{
+ \half (\log \CL)_{\phi_i \bar \phi_j}
(\p_{\mu} \phi_i) (\p^{\mu} \bar \phi_j)
- \half (\log \CL)_{\s_i \bar \s_j}
(\p_{\mu} \s_i) (\p^{\mu} \bar \s_j) +}
$$
+ \half \epsilon^{\mu \nu} \big(
(\log \CL)_{\phi_i \bar \s_j}
(\p_{\mu} \phi_i) (\p_{\nu} \bar \s_j)
+ (\log \CL)_{\s_j \bar \phi_i}
(\p_{\mu} \bar \phi_i) (\p_{\nu} \s_j) \big) \big]
$$
The subscripts on $\CL$ denote derivatives with respect
to the scalar fields, {\it e.g.} $(\log \CL)_{\phi_i \bar \phi_j}
= (\p^2 / \p \phi_i \p \bar \phi_j) \log \CL$.
Deriving \actcomp\ one may find helpful some formulas from appendix D
where we discuss in detail the component action of a free chiral superfield.
A careful reader may notice that \actcomp\ has the structure
reminiscent of $\CN=1$ supergravity in four dimensions.
In particular, it is convenient to introduce
the K\"ahler potential $K$:
\eqn\clk{ \CL = \exp (- K) }
so that the superspace action \generalact\ takes the form:
\eqn\generalkact{ \int d^2x d^4 \th E_0^{-1} e^{- K} =
\int d^2x d^4 \th E^{-1} e^{- 2V} e^{- K}.}
Performing the superspace integration, one finds
Lagrangian for the bosonic fields:
\eqn\actb{ L = e^{-2 \tilde \varphi} \big[
R^{(2)} + 4 (\p_{\mu} \tilde \varphi )
(\p^{\mu} \tilde \varphi )
- \half K_{\phi_i \bar \phi_j}
(\p_{\mu} \phi_i) (\p^{\mu} \bar \phi_j) + }
$$
+ \half K_{\s_i \bar \s_j}
(\p_{\mu} \s_i) (\p^{\mu} \bar \s_j)
- \half \epsilon^{\mu \nu} \big(
K_{\phi_i \bar \s_j}
(\p_{\mu} \phi_i) (\p_{\nu} \bar \s_j)
+ K_{\s_j \bar \phi_i}
(\p_{\mu} \bar \phi_i) (\p_{\nu} \s_j) \big) \big]
$$
where we introduced a new dilaton field
$\tilde \varphi = \varphi + \half K$
invariant under generalized K\"ahler transformations \ktransform:
\eqn\ktrans{ K \longrightarrow K
+ \Lambda_1 (\phi_i, \s_j )
+ \overline \Lambda_1 (\overline \phi_{\bar i},
\overline \s_j )
+ \Lambda_2 (\phi_i, \bar \s_j )
+ \overline \Lambda_2 (\bar \phi_{\bar i}, \s_j ) }
Under this K\"ahler transformation the original dilaton field
$\varphi$ is shifted in the opposite way, so that the linear
combination $\tilde \varphi = \varphi + \half K$ remains invariant.
Since the K\"ahler metric is invariant under \ktrans\ as well,
both the superspace action \generalkact\ and the corresponding
component action \actb\ are manifestly invariant under
the generalized K\"ahler transformations \ktrans.

Now we are in position to identify the function $K$
that would reproduce the effective action \act\ of Type IIA
theory on a Calabi-Yau four-fold.
Namely, the superspace action \generalkact\ gives the effective
action \act\ if $K$ is the total K\"ahler potential \ktot.
This form of the superspace action might be expected for
a number of reasons. First of all, it is similar to the
superspace action of $\CN=1$ supergravity in four dimensions.
Moreover, Type IIA supergravity on a Calabi-Yau four-fold has
a breathing mode corresponding to rescaling of the volume
$\CV \to c^2 \CV$ and simultaneous shift of the dilaton
$\varphi \to \varphi + \log c$, {\it cf.} \alagr.
Therefore, the superspace action is expected to be
a function of $2V + K$, where $K$ is the total K\"ahler
potential given by \ktot\ - \ksig.

To summarize, we constructed superspace Lagrangians
describing $\CN=(2,2)$ dilaton supergravity coupled
to matter superfields and found projection formulas
that allow one to rewrite integrals over the entire
superspace in terms of component fields.
Therefore, we provide a superspace formulation of
the effective field theories constructed from compactification
of Type IIA string theory on Calabi-Yau four-folds,
as well as more general $\CN=(2,2)$ sigma-models
with torsion coupled to dilaton supergravity, {\it cf.} \WGRN.

Incorporation of superpotential terms is more subtle.
These terms are superinvariants obtained by integration
only over a half of the superspace, {\it cf.} \superaction.
Unfortunately, unlike \cproj\ and \tproj, the chiral and
twisted chiral density projectors in $\CN=(2,2)$ dilaton
supergravity do not simply follow from the full superspace
projector \newproj.
However, by dimensional arguments and from an examination
of the index structure of the possible terms,
the chiral density projection formula must look like:
\eqn\cprojexp{ \int d^2x \int d^2 \th \CE^{-1} W
= \int d^2x e^{-1} \big[ \hat \nabla^2 + \ldots
- {1 \over 2} \overline H - \psi^{\dot -}_{\pp} \psi^{\dot +}_{\mm}
+ \psi^{\dot -}_{\mm} \psi^{\dot +}_{\pp} \big] W \vert }
where the dots stand for term containing $\la_{\a}$
or terms linear in covariant derivatives\foot{These
terms will not affect the action of bosonic fields.}.
By the similar reasoning, the twsited chiral density
projection formula must look like \tproj:
\eqn\tprojexp{ \int d^2x \int d^2 \th \tilde \CE^{-1}
\tilde W = \int d^2x e^{-1} \big[ \hat \nabla_{\dot -} \hat \nabla_{+}
+ \ldots - {1 \over 2} \bar G + \psi^{-}_{\pp} \psi^{\dot +}_{\mm}
+ \psi^{\dot +}_{\pp} \psi^{-}_{\mm} \big] \tilde W \vert }
Although normalization and coefficients in \cprojexp\
and \tprojexp\ may not be correct,
the terms quadratic in the gravitino are rather
general and, in particular, are independent on de-gauging.
So, we infer that one effect of the superpotential is
to produce a mass term for the gravitino fields \GGW:
\eqn\gravmass{m_{\psi_{\pp}^{\dot -}} \sim W,
\quad m_{\psi_{\pp}^{-}} \sim \tilde W}
It is this property of $\CN=(2,2)$ supergravity that was needed
in \refs{\GVW,\Gukov} in order to find the superpotentials
\wone\ and \wtwo\ induced by Ramond-Ramond fluxes.
Moreover, let us demonstrate that \cprojexp\ and \tprojexp\
lead to the expected structure of the scalar potential \scal.
Extending the computation of \actcomp, from \superaction\
we get the action of the auxiliary fields:
\eqn\auxact{ L_{\rm aux} \sim
\CL \big[ \vert \half H - i \bar A_i (\log \CL)_{\phi_i} \vert^2
+ \vert \half G - i \bar B_j (\log \CL)_{\s_j} \vert^2
+ (\log \CL)_{\phi_i \bar \phi_j} A_i \bar A_j - }
$$
- (\log \CL)_{\s_i \bar \s_j} B_i \bar B_j \big]
- i (W_{\phi_i} - (\log \CL)_{\phi_i}) A_i
- W ( \half \bar H + i A_i (\log \CL)_{\phi_i}) -
$$
$$
- i (\tilde W_{\s_i} - (\log \CL)_{\s_i}) B_i
- \tilde W ( \half \bar G + i B_i (\log \CL)_{\s_i})
+ {\rm c.c.}
$$
Integrating out the auxiliary fields and using \clk,
we find the expected scalar potential \scal:
$$
L_{\rm aux} \sim e^K \big(
K_{\phi_i \bar \phi_j}^{-1} (D_{\phi_i} W) (D_{\bar \phi_j} \bar W) -
K_{\s_i \bar \s_j}^{-1} (D_{\s_i} \tilde W) (D_{\bar \s_j} {\bar {\tilde W}})
- \vert W \vert^2 - \vert \tilde W \vert^2 \big)
$$
with the covariant derivatives \covd.


\newsec{Compactification of Type IIB String Theory on Calabi-Yau
Four-folds}

Compactification of Type IIB string theory on a Calabi-Yau
four-fold $X$ leads to a chiral $\CN=(0,4)$ supersymmetric
effective field theory in two non-compact dimensions. In the
low-energy limit this theory is described by $\CN=(0,4)$
supergravity coupled to scalar superfields. In this section
we perform a Kaluza-Klein reduction on a Calabi-Yau four-fold
$X$ and, in particular, find that the supergravity multiplet
includes a real scalar dilaton field instead of an $SU(2)$
gauge field \refs{\BS,\Nishino}. A manifestly supersymmetric
formulation of such $\CN=(0,4)$ supergravity theory will be
presented in the next section. In this section we show that
Kaluza-Klein harmonics combine into $\CN=(0,4)$ scalar superfields
\refs{\Nishino, \GatesRana}. Furthermore, since we deal with
$\CN=(0,4)$ supersymmetry, there is a significant difference
between right-movers and left-movers. Namely, all left-moving
modes are supersymmetry singlets. We will also see that the
difference between the zero-point energy of the left-movers
and the right-movers is proportional to the Euler number of $X$ \DM.

In the large volume limit the bosonic spectrum of light modes in
Type IIB string theory includes the metric $g_{MN}$, the dilaton
$\varphi$, the axion $l$, the 4-form tensor $D_{MNPQ}$ and two
tensor fields $B^{RR}_{MN}$ and $B^{NS}_{MN}$. Together with the
fermionic superpartners all these fields fit into Type IIB supergravity
multiplet. Non-perturbative Type IIB string theory is invariant
under $SL(2,\Z)$ duality group.  In order to see the action of this
group on the supergravity fields, it is convenient to define the
following quantities:
\eqn\ladef{\la = l +i e^{- \varphi}}
\eqn\matdef{{\bf M} = {1 \over \i \la}
\pmatrix{\vert \la \vert^2 & - \r \la \cr - \r \la & 1} }
\eqn\hfielddef{ {\bf H}_{MNP} = \pmatrix{ \p_{[M} B^{NS}_{NP]}
\cr ~\cr \p_{[M} B^{RR}_{NP]} } }
and
\eqn\ffielddef{ F_{MNPQR} = \p_{[M} D_{NPQR]}
+ {3 \over 4} B^{NS}_{[MN} \p_P B^{RR}_{QR]} }
Then, the field strength $F_{MNPQR}$ is a singlet under the $SL(2, \Z)$
duality group, while ${\bf H}$ transforms as a ``vector".  Finally,
$SL(2,\Z)$ acts on a complex scalar $\la$ in the usual way:
\eqn\sltwoz{\la \to {a \la + b \over c \la + d}}
where the integer numbers $a$, $b$, $c$ and $d$ satisfy $ad-bc=1$.

The five-form field strength $F$ is self-dual:
\eqn\selfdual{F = \ast F}
Although this equation can not be derived from any action, for a moment
we ignore this subtlety and write bosonic ``Type IIB supergravity
Lagrangian" simply as:
\eqn\blagr{ L_{(10)} = \sqrt{-g} \big[ - {1 \over 4} R^{(10)} +
{1 \over 16} \Tr \Big( \p {\bf M} \cdot \p {\bf M} \Big) +
{3 \over 16} {\bf H}^T \cdot {\bf M} \cdot {\bf H} +
{5 \over 24} F^2 + \ldots \big]}
where the dots stand for higher derivative terms.

As in Type IIA theory, in order to find the zero-mode spectrum we
have to expand Type IIB supergravity fields in harmonic $(p,q)$-forms
on the space $X$.  The metric modes are exactly the same as in
\defs. Namely, from the reduction of $g_{MN}$ we find $h^{1,1}$
real scalars $s_i$, $h^{3,1}$ complex scalars $\phi_j$ and the
two-dimensional metric $g_{\mu \nu}$. It turns out that one of
the scalars $s_i$ comes into the two-dimensional supergravity
multiplet.  Namely, it is the Kaluza-Klein mode corresponding to
the volume of the Calabi-Yau four-fold:
$$
\CV = d_{ijkl} s^i s^j s^k s^l
$$
where $d_{ijkl}$ are the intersection numbers of $X$ given by \dijkl.
With this mode excluded, the K\"ahler deformations of the metric
yield $h^{1,1}-1$ scalars $\check s_i = \CV^{-{1 \over 4}} s_i$
satisfying the condition:
$$
d_{ijkl} \check s^i \check s^j \check s^k \check s^l =1
$$

Expanding the doublet of tensor fields as:
\eqn\bees{B^{NS} = \sum_{i = 1}^{h^{1,1}} r_i \omega^{(1,1)}_i,
\quad \quad B^{RR} = \sum_{i = 1}^{h^{1,1}} t_i \omega^{(1,1)}_i, }
we get pairs of real scalars $r_i$ and $t_i$, $h^{1,1}$ in number.
All these modes are both right-moving and left-moving.

Expansion of the self-dual field $D$ is a bit subtle. Namely,
instead of $D$ one has to expand the field strength $F \sim
\sum_i \p_{\mu} u_i \cdot \omega_i^{(4)}$ and impose the self-duality
condition \selfdual.  Depending on whether the form $\omega_i^{(4)}
\in H^4 (X,\IR)$ is self-dual or anti-self-dual the scalar field
$u_i$ is left-moving or right-moving, respectively.  Therefore, we
have to distinguish carefully self-dual and anti-self-dual harmonics
of $F$.  To this end we recall some topological properties of Calabi-Yau
four-folds. There is a decomposition of the space of the middle
dimensional forms on $X$:
$$
H^4 (X, \IR) = B_+ (X) \oplus B_- (X)
$$
where we denote by $B_+ (X)$ (resp. $B_- (X)$) the space of
(anti-)self-dual 4-forms on $X$.   Let us call the corresponding
dimensions $b_{\pm} = {\rm dim} B_{\pm} (X)$.  Then $b_+$ and
$b_-$ are related by the Hirzebruch signature \foot{The explicit
form of the Pontryagin classes is given by:
$$
p_1 = - {1 \over 2} \tr R^2, \quad
p_2 = - {1 \over 4} \tr R^4 + {1 \over 8} (\tr R^2)^2
$$} (see {\it e.g.} \Klemm):
$$
\tau (Q) = b_+ - b_- = {1 \over 45} \int_X (7 p_2 - p_1^2) =
{ \chi \over 3} + 32
$$
of the quadratic form $Q(\omega_1, \omega_2) = \int_X \omega_1 \wedge
\omega_2$. On the other hand, we also have $b_4 = b_+ + b_- = 2 + 2
h^{3,1} + h^{2,2}$. Hence, using \htwotwo\ and \euler\ we find:
\eqn\bplus{b_+ = 47 + 3 h^{1,1} + 4 h^{3,1} - 2 h^{2,1}}
and
\eqn\bminus{b_- = -1 + h^{1,1} + 2 h^{3,1}}

Actually, we can be a little bit more precise.  All the forms of
Hodge type $(3,1)$ or $(1,3)$ are anti-self-dual, while the $(4,0)$-
and $(0,4)$-forms on a Calabi-Yau four-fold are self-dual \GVW.
Therefore, from \bplus\ and \bminus\ we find that:
\eqn\bmplus{b^{(2,2)}_+ = 45 + 3 h^{1,1} + 4 h^{3,1} - 2 h^{2,1}}
and
\eqn\bmminus{b^{(2,2)}_- = h^{1,1} -1}

Now we expand the self-dual field strength $F$ as:
\eqn\fdecomp{F =
\sum_{i=1}^{b^{2,2}_+} (\p_{\mu} u_i) \omega^{(+)}_i +
\sum_{j=1}^{b^{2,2}_-} (\p_{\mu} v_j) \omega^{(-)}_j +
\sum_{k=1}^{h^{3,1}} (\p_{\mu} p_k) \omega^{(3,1)}_k +
(\p_{\mu} q) \Omega + {\rm c.c.} }
where the scalar fields $u_i$ and $v_j$ are real, while $p_k$ and $q$
are complex. As we explained above, $u_i$ and $q$ must be left-moving,
$v_j$ and $p_k$ must be right-moving.  In particular, the former are
singlets with respect to four left supercharges satisfying:
$$
\{ Q^i_+, Q^j_+ \} = \delta^{ij} P_+
$$
Note that the Kaluza-Klein modes of the self-dual field $F$ associated
with $(2,1)$-harmonic forms give 2-form field strengths of massless
vector fields in two dimensions and, therefore, do not lead to new
propagating degrees of freedom.

Unlike $\CN=(2,2)$ theory constructed from compactification of Type
IIA string theory, in Type IIB compactification on on a Calabi-Yau
four-fold $X$ not all the fermionic modes can be determined by
$\CN=(0,4)$ supersymmetry. In the right sector we still can use
supersymmetry arguments to conclude that two-dimensional supergravity
multiplet contains $2 {\rm ind} (\D)$ fermions and $2 {\rm ind} (\D)$
Rarita-Schwinger fields that come from the corresponding spin-${1
\over 2}$ and spin-${3 \over 2}$ fields in Type IIB supergravity.
On a Calabi-Yau four-fold the Dirac index is given by:
\eqn\dindex{{\rm ind} (\D) = {1 \over 1440} \int_X
\Big( {7 \over 4} p_1^2 - p_2 \Big) = 2}
in accordance with $\CN=(0,4)$ supersymmetry. Furthermore, all the
right-moving scalars found above ($\phi_i$, $\bar \phi_{\bar i}$,
$p_i$, $\bar p_{\bar i}$, $\check s_j$, $r_j$, $t_j$, $v_j$, $\varphi$
and $l$) are accompanied by right-moving fermions.  Simple counting
gives:
\eqn\rferm{n_+ = 4 h^{3,1} + 4 h^{1,1}}
for the total number of the right-moving fermions \foot{Note that $n_+$
is divisible by 4.}. There are also left-moving fermions which are
supersymmetry singlets. The number of left-moving fermions, however, is
not determined by supersymmetry. So, it has to be computed separately.
Since the fermions in question come from the Type IIB gravitinos, their
number (minus the number of right-moving fermions) is given by the
Rarita-Schwinger index:
$$
n_- - n_+ = 2 {\rm ind} (\D_{3/2})
$$
Using \rferm\ and the explicit expression for the Rarita-Schwinger index
on a Calabi-Yau four-fold $X$:
\eqn\rsindex{{\rm ind} (\D_{3/2}) = {1 \over 180} \int_X
\Big( {37 \over 4} p_1^2 - 31 p_2 \Big)
= - 4h^{1,1} - 4h^{3,1} + 4h^{2,1}}
we obtain:
\eqn\lferm{n_- = 4 h^{2,1}}

Now we are ready to assemble the supermultiplets.  Combining the
left-moving bosonic modes with the fermion fields we get the following
supermultiplets:
$$
{\rm a~ gravitational~ multiplet:}
\quad g_{\mu \nu},
\quad \CV
$$
$$
h^{3,1}~ {\rm scalar~ multiplets~} \Phi_i:
\quad \phi_i,
\quad \bar \phi_i,
\quad p_i,
\quad \bar p_i
$$
$$
h^{1,1}~ {\rm scalar~ multiplets~} \S_j:
\quad \check s_j,
\quad r_j,
\quad t_j,
\quad v_j,
\quad \varphi,
\quad l
$$
Performing a reduction of the ten-dimensional supersymmetry
conditions one can easily check that these fields indeed represent
bosonic components of the supermultiplets as stated.  Note,
all the matter multiplets include four real scalar fields in
accordance with the general classification of scalar superfields
\refs{\BS, \HPT, \Nishino, \GatesRana}. However, the content of
the gravitational multiplet is different from what was usually
studied in $\CN=(0,4)$ supergravity theories \refs{\BS, \Nishino}.
In compactification of Type IIB string theory on a Calabi-Yau
four-fold we find that the gravitational multiplet includes a real
scalar $\CV$ instead of $SU(2)$ gauge field.  In the next section we
describe the superspace formulation of this $\CN=(0,4)$ supergravity
using the Goldstone approach.

In order to find the low-energy effective action one has to substitute
\defs, \bees\ and \fdecomp\ into \blagr. Integrating over the internal
space by means of the formulas \phimet, \smet\
we get the following two-dimensional action for bosonic fields:
$$
L_{(2)} = \sqrt{-g} \CV \big[ - {1 \over 4} R^{(2)} +
{1 \over 2} (\p_+ \la) (\p_- \la)
+ {1 \over 2} \CG_{\phi_i \bar \phi_{\bar j}}
(\p_+ \phi^i) (\p_- \bar \phi^{\bar j})
+ {1 \over 2} \CG_{\phi_i \bar \phi_{\bar j}}
(\p_0 p^i) (\p_+ \bar p^{\bar j}) +
$$
$$
+ {1 \over 2} \CG_{\s_i \bar \s_j}
\big( (\p_+ s^i) (\p_- s^{j}) +
(\p_+ r^i) (\p_- r^{j}) +
(\p_+ t^i) (\p_- t^{j}) +
(\p_0 v^i) (\p_+ v^{j}) \big) +
$$
\eqn\bact{ + {1 \over 2} Q_{ij} (\p_0 u^i) (\p_- u^j)
+ {1 \over 2} (\p_0 q) (\p_- q) + {\rm c.c.} \big] }
This Lagrangian describes non-linear sigma-model interacting with
$\CN=(0,4)$ supergravity.   The target space of the left-moving
fields is the cotangent bundle to the moduli space of the Calabi-Yau
manifold
$X$,
$T^* \CM_c (X) \times T^* \CM_{\CK} (X)$, {\it cf.} \Nishino.
Since the moduli space itself is a K\"ahler space, this result
agrees with the general analysis of $\CN=(0,4)$ supersymmetric
sigma-models. According to \HPT, $\CN=(0,4)$ supersymmetric
sigma-model is based on a target space which has three covariantly
constant (with respect to $\nabla_+$) complex structures which obey
the quaternionic algebra:
$$
J_r J_s = - \delta_{rs} + {f_{rs}}^t J_t
$$

Another interesting feature that we expect to see in this $\CN=(0,4)$
theory is $SL(2,\IR)$ symmetry of classical Type IIB supergravity.
Apart from $\varphi$, $l$, $r_i$ and $t_i$, all the scalar fields
listed above are singlets with respect to this symmetry. The
complex scalar $\la = l +i e^{- \varphi}$ transforms as \sltwoz\ under
$SL(2,\IR)$ duality transformation, with $a$, $b$, $c$ and $d$ real
numbers obeying $ad-bc=1$. The doublet of real scalar fields $(r_i
,t_i )$ transforms as a vector under the general $SL(2, \IR)$
transformation. In other words, only the scalar multiplets $\S_i$
transform non-trivially under this symmetry, while all the other
fields, including the supergravity itself, are $SL(2, \IR)$-singlets.

Finally, we remark that from T-duality with Type IIA string theory
on a Calabi-Yau four-fold we expect an anomaly similar to \sethrel\
in Type IIB compactification on a Calabi-Yau four-fold. Recall that
due to the global anomaly \sethrel, in Type IIA vacuum we had to
include $N = {\chi \over 24}$ fundamental strings filling two-dimensional
space-time to cancel the tadpole. Under a T-duality in one of the
space-time directions the winding modes of these strings transform
into ${\chi \over 24}$ momentum modes:
\eqn\momentum{ P_+ - P_- = {\chi \over 24}}
in the Type IIB vacuum corresponding to compactification on the same
Calabi-Yau four-fold $X$. Here, for the sake of simplicity, we assumed
that there are no background fluxes.  One can interpret \momentum\ as
the difference in the zero-point energy of the left-moving and the
right-moving Kaluza-Klein modes \DM. In order to see this, we note that
a free boson on a circle has vacuum energy $- {1 \over 24}$ and a periodic
fermion has vacuum energy $+ {1 \over 24}$.  Therefore, due to $\CN=(0,4)$
supersymmetry, in the right sector bosonic and fermionic contributions
cancel each other, {\it i.e.} $P_+=0$. In the left sector we have $48 +
6h^{1,1} + 6h^{3,1} - 2h^{2,1}$ bosonic modes corresponding to the fields
$\phi_i$, $\bar \phi_{\bar i}$, $\check s_j$, $r_j$, $t_j$, $u_i$,
$q$, $\bar q$, $\varphi$ and $l$ along with $n_- = 4h^{1,1}$ fermionic
modes, {\it cf.} \lferm. Hence, the total vacuum momentum in the left
sector is non-zero and is given by the following formula:
\eqn\lmomentum{P_- = {1 \over 24} (48 + 6h^{1,1} + 6h^{3,1} - 6h^{2,1})}
Using $P_+ = 0$ and the explicit expression \euler\ for the Euler
number, one can easily obtain the formula \momentum.


\newsec{Superspace Formulation of $\CN=(0,4)$ Dilaton Supergravity}

In this section we construct $\CN=(0,4)$ dilaton supergravity that
arise, for example, in Type IIB superstring compactification on
Calabi-Yau four-folds.  As we demonstrated in the previous section
such a theory has a number of distinct features which are absent
in the existing superspace formulations of two-dimensional $\CN=
(0,4)$ supergravities.  Namely, unlike the standard formulations
with gauged $SU(2)$ $R$-symmetry \refs{\BS, \Nishino},
new supergravity does not have a gauged symmetry and
the supergravity multiplet contains a real dilaton field $\CV$.
Below we present a superspace construction of this theory
obtained via de-gauging $SU(2)$ symmetry.

First let us remind that gauged $\CN=(0,4)$ supergravity is defined
in superspace by the following set of constraints \DGR:
$$
[ \nabla_{+ i} , \nabla_{+ j} \} = 0, \quad
[ \nabla_{+ i} , \nabla_{\dot +} {}^j \} = i 2 \d_i^j \nabla_{\pp} \quad ,
$$
\eqn\coldsugra{[ \nabla_{+ i} , \nabla_{\pp} \} = 0 \quad , \quad
[ \nabla_{+ i} , \nabla_{\mm} \} =
- i [ \bar \S^+ {}_i \CX - \bar \S^+ {}_j \CY_i {}^j ] \quad ,}
$$
[ \nabla_{\pp} , \nabla_{\mm} \} = - \half [
\S^{+ i} \nabla_{+ i} + \bar \S^+ {}_i \nabla_{\dot +} {}^i
+ \CR \CX + i \CF_i {}^j \CY_j {}^i ]
$$
on the covariant derivatives
$\nabla_{A} \equiv ( \nabla_{+ i}, \nabla_{\dot +} {}^i, \nabla_{\pp},
\nabla_{\mm} )$:
\eqn\ccderiv{ \nabla_{A} = E_{A} {}^{B} D_{B} + \Lambda_{A} \CX +
i \CA_{A i} {}^j \CY_j {}^i .}
Here ${E_{A}}^{B}$ is the supervielbein, and $\CX$ and $\CY^i_j$
are the Lorentz and $SU(2)$ symmetry generators, respectively.
The superfield $\CA_{A i} {}^j$ is $SU(2)$ gauge connection, while
$\Lambda_{A}$ stands for the Lorentz spin-connection. We write $[~,~\}$
for the graded (anti-)commutator. Finally, $D_{A}$ denotes the flat
space fermi and bose derivatives $D_{A} \equiv (  \bar D_{+ i}, D_+
{}^i, \p_{\pp}, \p_{\mm}$).

The Lorentz generators act on $\nabla_{A}$ as follows:
$$
[ \CX , \nabla_{+ i} \} = {1 \over 2} \nabla_{+ i} \quad , \quad
[ \CX , \nabla_{\dot +} {}^{i} \} = {1 \over 2} \nabla_{\dot +}{}^{i}
\quad ,
$$
$$
[ \CX , \nabla_{\pp} \} = \nabla_{\pp} \quad , \quad
[ \CX , \nabla_{\mm} \} = - \nabla_{\mm} \quad.
$$

Similarly, for the action of $SU(2)$ gauge symmetry generators we have:
$$
[ \CY_j {}^k , \nabla_{+ i} \} = \d_i^k \nabla_{+ j} -
\half \d_j^k \nabla_{+ i} \quad ,
$$
$$
[ \CY_j {}^k , \nabla_{\dot +} {}^i \} = - \d_j^i \nabla_{\dot +} {}^k
+ \half \d_j^k \nabla_{\dot +} {}^ i \quad ,
$$
$$
[ \CY_j {}^k , \nabla_{\pp} \} = 0 \quad , \quad
[ \CY_j {}^k , \nabla_{\mm} \} = 0 \quad .
$$

The constraints \coldsugra\ lead to a set of Bianchi identities
that are solved if:
$$
\nabla_{\dot +} {}^i \bar \S^{+ j} = 0 \quad , \quad
\nabla_{+ i} \S^{+ j} = \half \d_i^j \CR + i \CF_i {}^j \quad ,
$$
$$
\nabla_{+ i} \CR = i 2 \nabla_{\pp} \bar \S^+ {}_i \quad , \quad
\nabla_{+ i} \CF_j {}^k = - 2 \d_i^k \nabla_{\pp} \bar \S^+ {}_j
+ \d_j^k \nabla_{\pp} \bar \S^+ {}_i \quad .
$$

The first step in obtaining two-dimensional $\CN=(0,4)$ supergravity
theory that does not contain a gauged $SU(2)$ is to note that the
covariant derivative in \ccderiv\ can be split as:
\eqn\sepderiv{ \nabla_{A} = \hat \nabla_{A} + i  \CA_{A k}{}^l
\CY_l{}^k .}
Since the superspace covariant derivative $\hat \nabla_{A}$
does not contain the $SU(2)$ connection nor the generator,
it can not describe two-dimensional $\CN=(0,4)$ supergravity
with gauged $SU(2)$ symmetry. We next use \sepderiv\ to derive
the form of the commutator algebra for the $\hat \nabla_{A}$
operators.

A straightforward set of calculations leads to:
$$
[ \hat \nabla_{+ i} , \hat \nabla_{+ j} \} = -i
[~  {\cal A}_{+  i \, j} {}^k ~+~  {\cal A}_{+ j \, i} {}^k ~]
 \hat \nabla_{+ k}  \quad ,
$$
$$
[ \hat \nabla_{+ i} , \hat \nabla_{\dot +} {}^j \} = i 2 \d_i^j \hat
\nabla_{\pp} + i  {\bar {\cal A}}{}_+ {}^j {}_i {}^k \, \hat
\nabla_{+ k} + i {\cal A}_{+  i \, k} {}^j \,
\hat \nabla_{\dot +} {}^k \quad ,
$$
\eqn\newcsugra{[ \hat \nabla_{+ i} , \hat \nabla_{\pp} \} =
i \CA_{\pp i} {}^k \hat \nabla_{+ k} \quad , \quad
[ \hat \nabla_{+ i} , \hat \nabla_{\mm} \} =
i \CA_{\mm i} {}^k \hat \nabla_{+ k} - i \S^+ {}_i \CX \quad ,}
$$
[ \hat \nabla_{\pp} , \hat \nabla_{\mm} \} = - \half [\S^{+ i}
\hat \nabla_{+ i} + \bar \S^+ {}_i \hat \nabla_{\dot +} {}^i + \CR \CX]
$$
where the connection superfields now explicitly appear on the right-hand
side of the equations. The leading component of $ {\cal A}_{+ i \, j}{}^k$
is a component of the gauged supergravity multiplet that could be
eliminated by a gauge transformation in the $SU(2)$ gauge-symmetric
phase.

At this stage, we have completed half of the de-gauging process.  The
second half consists of specifying the spinorial $SU(2)$ connections
in terms of some components of another (matter) multiplet that is
consistent with the two-dimensional $\CN=(0,4)$ supergravity theory.
For this purpose, we introduce the second of the four distinct
$\CN=(0,4)$ scalar multiplets (SM-II) that were discussed in
\refs{\Nishino, \GatesRana}.  In the case of rigid supersymmetry this
multiplet is described by:
$$
D_{+ i} \CV = i \la^- {}_i \quad , \quad \bar \CV
= \CV \quad , \quad \varphi^i {}_i = 0 \quad ,
$$
$$
D_{+ i} \varphi_j {}^k = 2 \d_i^k \la^- {}_j - \d_j^k \la^- {}_i \quad ,
\quad \bar \varphi_j {}^i = \varphi_i {}^j \quad ,
$$
$$
D_{\dot +} {}^i \la^- {}_j = \d_j^i \p_{\pp} \CV
+ i \p_{\pp} \varphi_j {}^i \quad , \quad
D_{+ i} \la^- {}_j = 0 \quad .
$$
In the locally supersymmetric theory one has to replace
$D_{\a}$ by $\nabla_{\a}$.

The triplet of spin-$0$ fields $\varphi_j {}^i$ can be ``eaten'' by
the triplet of spin-$1$ fields in the minimal $\CN=(0,4)$ supergravity
multiplet and thus become their longitudinal components via the usual
Goldstone mechanism. This eliminates the local $SU(2)$ symmetry.  The
scalar and spinor field of the matter multiplet become the dilaton
and dilatino.

We are thus led to conjecture that the new form of two-dimensional
$\CN=(0,4)$ supergravity with a component spectrum given by:

${~~~~~~~~}$ $e_{\mu}{}^{\nu}  $ ${~~~~}$graviton,

${~~~~~~~~}$ $\psi_{\mu}{}^{+ i}$ ${~}$ $SU(2)$-doublet, gravitino,

${~~~~~~~~}$ $A_{\mu i}{}^j $ ${~~}$ $SU(2)$-triplet, vector auxiliary
fields,

${~~~~~~~~}$ $\la^{-}{}_i $  ${~~~}$ $SU(2)$-doublet, dilatino field and

${~~~~~~~~}$ $\CV$ ${~~~~~~~}$real dilaton field

\noindent
may be constructed with \newcsugra\ as its starting point.  We note
that the chirality of the dilatino is opposite to that of the gravitino.

In order to gain a control over the component field content
of the  theory, we must impose the following constraints:
$$
\CA_{+ ij} {}^k = [ 2 \la^- {}_j \d^k_i - \la^- {}_i \d^k_j ] \quad ,
$$
$$
\bar \CA_+ {}^i {}_j {}^k =
- [2 \bar \la^- {}_j \d^k_i - \bar \la^- {}_i \d^k_j] \quad .
$$
At lowest order in $\theta^{\a}$, the field $\CA_{+ ij}{}^k$ is
a component field that is absent in the $SU(2)$ gauge-symmetric
phase of the theory; it can be set to zero in the Wess-Zumino gauge.
On the other hand, when the $SU(2)$ symmetry is broken, part of
this field becomes dynamical. In general, $\CA_{+ ij}{}^k$
contains $SU(2)$ representations of spin-${1 \over 2}$ and spin-${3
\over 2}$.  However, the above constraints eliminate the pure spin-${3
\over 2}$ representation of $SU(2)$.

With this result substituted into \newcsugra,
$$
[ \hat \nabla_{+ i} , \hat \nabla_{+ j} \} = -i
[ \la^- {}_i \hat \nabla_{+ j} + \la^- {}_j \hat \nabla_{+ i} ] \quad ,
$$
$$
[ \hat \nabla_{+ i} , \hat \nabla_{\dot +} {}^j \} =
i 2 \d_i^j \hat \nabla_{\pp} + i [ 2 \bar \la^{- k} \d_i^j -
\bar \la^{- j} \d_i^k ] \hat \nabla_{+ k} +
i [ 2 \la^- {}_k \d_i^j - \la^- {}_i \d_k^j ]
\hat \nabla_{\dot +} {}^k \quad ,
$$
$$
[ \hat \nabla_{+ i} , \hat \nabla_{\pp} \} =
i \CA_{\pp i} {}^k \hat \nabla_{+ k} \quad ,
\quad [ \hat \nabla_{+ i} , \hat \nabla_{\mm} \} =
i \CA_{\mm i} {}^k \hat \nabla_{+ k} - i \S^+ {}_i \CX \quad ,
$$
$$
[ \hat \nabla_{\pp} , \hat \nabla_{\mm} \} =
- \half [\S^{+ i} \hat \nabla_{+ i} +
\bar \S^+ {}_i \hat \nabla_{\dot +} {}^i + \CR \CX]
$$
we can calculate the Bianchi identities:
\eqn\bianchi{ [ [ \hat \nabla_{+ i},
\hat \nabla_{+ j} \} , \hat \nabla_{+ k} \} +
[ [ \hat \nabla_{+ k}, \hat \nabla_{+ i} \} , \hat \nabla_{+ j} \} +
[ [ \hat \nabla_{+ j}, \hat \nabla_{+ k} \} , \hat \nabla_{+ i} \} = 0 }
$$
[[ \hat \nabla_{+ i} , \hat \nabla_{+ j} \} ,
\hat \nabla_{\dot +} {}^k \}
+ [[ \hat \nabla_{+ i} , \hat \nabla_{\dot +} {}^k \} ,
\hat \nabla_{+ j} \}
+ [[ \hat \nabla_{+ j} , \hat \nabla_{\dot +} {}^k \} ,
\hat \nabla_{+ i} \} = 0
$$
These will be satisfied if:
$$
\hat \nabla_{+ i} \la^- {}_j = - \la^- {}_i \la^- {}_j \quad ,
$$
$$
\hat \nabla_{+ i} \bar \la^{- j} =
i \la^- {}_i \bar \la^{- j} - i \CA_{\pp i} {}^j
+ \d_j^i \hat \nabla_{\pp} \CV \quad , etc.
$$

Let us now briefly comment on density projectors
in the new $\CN=(0,4)$ dilaton supergravity theory.
For a general superspace Lagrangian $\CL$, the component action
of $\CN=(0,4)$ gauged supergravity can be obtained by means of
the following projection formula:
$$
\int d^2 x d^2 \theta^{\pp} d^2 \bar \theta^{\pp} E^{-1} \CL =
\half \int  d^2 x d^2 \theta^{\pp} \CE^{-1} \Big[ \half C_{ij}
\hat \nabla_{\dot +} {}^i \hat \nabla_{\dot +} {}^j \Big] \CL \vert +
$$
\eqn\oldproj{ + \half \int d^2 x d^2 \bar \theta^{\pp} \bar
\CE^{-1} \Big[ \half C^{ij} \hat \nabla_{+ i} \hat \nabla_{+ j}
\Big] \CL \vert }
where the corresponding chiral and anti-chiral density
projector formulas look like:
$$
\int  d^2 x d^2 \theta^{\pp} \CE^{-1} \CL \vert =
i \int d^2 x \Big[ \half e^{-1} C^{ij} (\hat \nabla_{+ i} +
i4e \bar \psi_{\pp} {}^+ {}_i ) \Big] \hat \nabla_{+ j} \CL \vert
$$
$$
\int  d^2 x d^2 \bar \theta^{\pp} \bar \CE^{-1} \CL \vert =
i \int d^2 x \Big[ \half e^{-1} C_{ij} (\hat \nabla_{\dot +} {}^i +
i4e \psi_{\pp} {}^{+ i} ) \Big] \hat \nabla_{\dot +} {}^j \CL \vert
$$
Similar formulas also hold in the new two-dimensional
$\CN=(0,4)$ dilaton supergravity. The explicit expressions
for the density projectors can be obtained by a straightforward
but tedious computation substituting \sepderiv\ into \oldproj.


\vskip 30pt
\centerline{\it ``I found a way to make it work."}
\centerline{~~~~~~~~~~~~~~~~~~~~~~~~~~~~
~~~~~~~~~~~~~~~~~~~~~~~~~~~~~~~~~~~~~~~~Stanislaw Ulam}

\vskip 20pt
\centerline{\bf Acknowledgments}

We are grateful to Marc Grisaru, Martin Ro${\check {\rm c}}$ek
and John H. Schwarz for useful discussions. 
The research of S.J.G. is supported by the NSF grant No PHY-98-02551;
S.G. is supported in part by the Caltech Discovery Fund,
grant RFBR No 98-02-16575 and Russian President's grant No 96-15-96939.
The work of E.W. is supported in part by NSF Grant PHY-9513835
and the Caltech Discovery Fund.

\appendix{A}{World-Sheet Calculation of Type IIA String Amplitudes}

Consider compactification of Type IIA string theory on a Calabi-Yau
four-fold $X$.  Let us further assume that there are no background
Ramond-Ramond fluxes, so that two-dimensional space-time is flat.
{}From the world-sheet viewpoint, this compactification corresponds to
adjoining $c=(12,12)$ $\CN=(2,2)$ superconformal field theory (SCFT)
to free conformal theory with central charge $c=(3,3)$ that is
responsible for the two-dimensional space-time. It is the first part
that will be interesting to us. Namely, we are going to show that
two-point correlation function of vertex operators corresponding to
chiral and twisted chiral superfields is zero, {\it i.e.} that the
Zamolodchikov metric on the moduli space of Calabi-Yau four-folds is
block diagonal \split.

$\CN=(2,2)$ superconformal algebra consists of two $\CN=2$
supervirasoro algebras --- one left-moving and one right-moving
--- each generated by an energy-momentum tensor $T$, a current $J$
and two weight $3/2$ supercurrents $G^{\pm}$ with $J$-charge $Q= \pm
1$.  Recall that in a Kaluza-Klein reduction two-dimensional chiral
superfields come from harmonic $(3,1)$-forms on $X$, while twisted
chiral superfields correspond to harmonic $(1,1)$-forms.  Similar to
the three-fold case, we identify these fields with marginal operators
in $(c,c)$ and $(a,c)$ multiplets, respectively.  Let us call this
operators $\Phi_{(1,1)}$ and $\Phi_{(-1,1)}$.  They are neutral, $(Q,
\bar Q) = (0,0)$, and have conformal weight $1/2$. The lowest components
of (anti-)chiral multiplets must satisfy $2h = Q$ and $2 \bar h = \bar
Q$, so $\Phi_{(1,1)}$ and $\Phi_{(-1,1)}$ are not the lowest components
in the corresponding multiplets.  They can be obtained in the operator
product expansion of the supercurrents with operators $\Psi$:
\eqn\gpsiope{ 2 G^{\mp} (w, \bar w)
\cdot \Psi_{(\pm 1,1)} (z, \bar z) =
{1 \over w-z} \Phi_{(\pm 1, 1)} (z, \bar z)
+ {\rm reg} }
where ``reg" stands for the regular part.  Another operator product
expansion that will be useful to us is the following:
\eqn\gphiope{ 2 G^{\mp} (w, \bar w) \cdot
\Phi_{(\pm 1,1)} (z, \bar z) = {\rm reg} }

Now we are ready to demonstrate \split.  Consider a matrix element
of the target space metric $\CG_{\phi_i \bar \s_j}$ that mixes $(1,1)$
and $(3,1)$ moduli:
\eqn\opesplit{ {\CG_{\phi_i \bar \s_j} \over
\vert z - z' \vert^4 }
= \langle \Phi^{\phi_i}_{(1,1)} (z, \bar z) \cdot
\Phi^{\bar \s_j}_{(1,-1)} (z', \bar z') \rangle = }
$$
= \oint {dw \over 2 \pi i} \langle 2 G^{-} (w, \bar w)
\cdot \Psi^{\phi_i}_{(1,1)} (z, \bar z) \cdot
\Phi^{\bar \s_j}_{(1,-1)} (z', \bar z') \rangle = 0
$$
In the last equality we used the fact \gphiope\ that the operator
product of $G^{-} (w, \bar w)$ and $\Phi^{\bar \s_j}_{(1,-1)}
(z', \bar z')$ has no singularity as $w \to z'$.

One might think that $\CG_{\phi_i \s_j}$ would be non-zero since the
corresponding OPE has a singular part:
$$
2 G^{\pm} (w, \bar w) \cdot \Phi_{(\pm 1,1)} (z, \bar z) =
{\p \over \p z}
\Big( { \Psi_{(\pm 1, 1)} (z, \bar z) \over w-z} \Big)
+ {\rm reg}
$$
However, repeating the above arguments in the right sector one can
easily see that $\CG_{\phi_i \s_j}$ is also zero.  In fact, OPE is
singular in both left and right sectors only for $\langle
\Phi^{\phi_i}_{(1,1)}
\cdot
\bar \Phi^{\bar \phi_{\bar j}}_{(-1,-1)} \rangle$ and
$\langle \Phi^{\s_i}_{(-1,1)} \cdot
\bar \Phi^{\bar \s_j}_{(1,-1)} \rangle$
which correspond to the metric $\CG_{\phi_i \bar \phi_{\bar j}}$
for chiral multiplets and the metric $\CG_{\s_i \bar \s_j}$ for
the twisted chiral multiplets, respectively. Therefore, we conclude
that the target space is locally a product of the manifold $\CM_c
(X)$ parametrized by the chiral fields $\phi_i$ and the manifold
$\CM_{\CK} (X)$ spanned by the twisted chiral fields $\s_j$.

\appendix{B}{Extra derivative constraints arising from
de-gauging $\CN=(2,2)$ non-minimal supergravity}

In this appendix we collect some more technical formulas
that arise in the construction of $\CN=(2,2)$ dilaton supergravity
via de-gauging $U(1)_A \otimes U(1)_V$ non-minimal supergravity.
Consistency of the de-gauging procedure requires that all terms
in \sugra\ with gauge symmetry generators $\CY$  and $\CY'$ vanish,
so that the commutator algebra of the new covariant derivative
$\hat \nabla_{\a}$ has the form \newsugra, {\it e.g.}:
$$
\{ \hat \nabla_+ , \hat \nabla_+ \} =
\{ \nabla_+ , \nabla_+ \}
- 2 \{ \nabla_+ , \la_+ \CY + \tilde \la_+ \CY' \} =
$$
$$
= -2 (\nabla_+ \la_+) \CY - 2 (\nabla_+ \tilde \la_+) \CY'
+ i \la_+ \nabla_+ + i \tilde \la_+ \nabla_+ =
$$
$$
= i (\la_+ + \tilde \la_+) (\hat \nabla_+ + \la_+ \CY + \tilde \la_+ \CY')
- 2(\nabla_+ \la_+ ) \CY -  2(\nabla_+ \tilde \la_+ ) \CY' =
$$
$$
= i (\la_+ + \tilde \la_+) \hat \nabla_+
+ [ -2 (\nabla_+ \la_+ ) + i \tilde \la_+ \la_+] \CY
+ [ -2 (\nabla_+ \tilde \la_+ ) + i \la_+ \tilde \la_+] \CY' =
$$
$$
= i (\la_+ + \tilde \la_+) \hat \nabla_+
+ [ -2 (\hat \nabla_+ \la_+ ) + i \tilde \la_+ \la_+
+ i \tilde \la_+ \la_+] \CY
+ [ -2 (\hat \nabla_+ \tilde \la_+ ) + i \la_+ \tilde \la_+
+ i \la_+ \tilde \la_+] \CY' =
$$
$$
= i (\la_+ + \tilde \la_+) \hat \nabla_+
+ [ -2 (\hat \nabla_+ \la_+ ) + 2i \tilde \la_+ \la_+] \CY
+ [ -2 (\hat \nabla_+ \tilde \la_+ ) + 2i \la_+ \tilde \la_+] \CY'
$$

Therefore, we obtain the following commutation relation:
$$
\{ \hat \nabla_+ , \hat \nabla_+ \} =
i (\la_+ + \tilde \la_+) \hat \nabla_+
$$
plus two constraints (Jacobi identities):
$$
(\hat \nabla_+ \la_+ ) - i \tilde \la_+ \la_+ =0 \quad , \quad
(\hat \nabla_+ \tilde \la_+ ) - i \la_+ \tilde \la_+ =0.
$$

Similar calculations lead to the following set of constraints:
$$
\hat \nabla_{+} \la_{+} - i \tilde \la_{+} \la_{+} = 0,
\quad , \quad
\hat \nabla_{+} \tilde \la_{+} - i \la_{+} \tilde \la_{+} = 0,
$$
$$
\hat \nabla_{-} \la_{-} + i \tilde \la_{-} \la_{-} = 0,
\quad , \quad
\hat \nabla_{-} \tilde \la_{-} - i \la_{-} \tilde \la_{-} = 0,
$$
$$
\hat \nabla_{+} \la_{-} + \hat \nabla_{-} \la_{+} +
i (\tilde \la_{+} \la_{-} - \tilde \la_{-} \la_{+}) =0,
$$
$$
\hat \nabla_{+} \tilde \la_{\dot -} + \hat \nabla_{\dot -} \tilde \la_{+} +
i (\la_{+} \tilde \la_{\dot -} -\la_{\dot -} \tilde \la_{+}) =0,
$$
$$
\hat \nabla_{+} \tilde \la_{-} + \hat \nabla_{-} \tilde \la_{+} +
2 i \tilde \la_{+} \tilde \la_{-} -
i  (\la_{+} \tilde \la_{-} + \la_{-} \tilde \la_{+})
- {i \over 2} \bar R =0,
$$
$$
\hat \nabla_{+} \la_{\dot -} + \hat \nabla_{\dot -} \la_{+} +
2 i \la_{+} \la_{\dot -} -
i (\tilde \la_{+} \la_{\dot -} + \tilde \la_{\dot -} \la_{+})
- {i \over 2} \bar F =0,
$$
$$
\tilde \la_{\pp} = -i (\hat \nabla_{+} \tilde \la_{\dot +} +
\hat \nabla_{\dot +} \tilde \la_{+}) + 2 \tilde \la_{+} \tilde \la_{\dot +}
+ (\la_{+} \tilde \la_{\dot +} + \tilde \la_{+} \la_{\dot +}),
$$
$$
\la_{\pp} = -i (\hat \nabla_{+} \la_{\dot +} +
\hat \nabla_{\dot +} \la_{+}) + 2 \la_{+} \la_{\dot +}
+ (\la_{+} \tilde \la_{\dot +} + \tilde \la_{+} \la_{\dot +}),
$$
$$
\tilde \la_{\mm} = -i  (\hat \nabla_{-} \tilde \la_{\dot -} +
\hat \nabla_{\dot -} \tilde \la_{-}) - 2 \tilde \la_{-} \tilde \la_{\dot -}
+ (\la_{-} \tilde \la_{\dot -} + \tilde \la_{-} \la_{\dot -}),
$$
$$
\la_{\mm} = -i  (\hat \nabla_{-} \la_{\dot -} +
\hat \nabla_{\dot -} \la_{-}) + 2 \la_{-} \la_{\dot -}
- (\la_{-} \tilde \la_{\dot -} + \tilde \la_{-} \la_{\dot -})
$$
By virtue of the above relations, the Jacobi identities \jacobi\
are automatically satisfied.
Furthermore, only half of the spinor fields are independent.
In section 4 we disscuss two natural solutions to these constraints:
when either $\eta_{\a}$ or $\tilde \eta_{\a}$ are put to zero.
In both cases the remaining spinor superfields can be expressed
in terms of an unconstrained real superfield.

\appendix{C}{Components of covariant derivatives
in $\CN=(2,2)$ dilaton supergravity}

The expressions for the covariant derivatives in $\CN=(2,2)$
dilaton supergravity evaluated at $\theta=0$ are, {\it cf.} \GGRS:
$$
\hat \nabla_{\a} \vert = \p_{\a}
$$
\eqn\ncomp{ \hat \nabla_{\mu} \vert =
\DD_{\mu} + \psi_{\mu}^{\a} \hat \nabla_{\a} \vert
+ \psi_{\mu}^{\dot \a} \hat \nabla_{\dot \a} \vert = }
$$
= \DD_{\mu} + \psi_{\mu}^{\a} \p_{\a} + \psi_{\mu}^{\dot \a} \p_{\dot \a}
$$
where $\DD_{\mu}$ is the fully covariant gravitational
derivative with the Lorentz connection
$\omega_{\mu} = \Lambda_{\mu} \vert$ that includes,
in addition to the ordinary connection, extra terms that
are bilinear in the gravitini $\psi_{\mu}^{\a}$, $\psi_{\mu}^{\dot \a}$.
Specifically, $\DD_{\mu}$ is defined to be \foot{In the notations
of \GGW\ this would correspond to
$\omega_{\mu} = \gamma_{\mu} = \varphi_{\mu}$.}:
$$
\DD_{\mu} = e_{\mu} + \omega_{\mu} \CX
$$

We also need expressions for the higher $\theta$ components
of the covariant derivatives. The $\theta^{\a}$ component
of $\hat \nabla_{\b}$ is defined by \GGRS:
\eqn\spincomp{ \hat \nabla_{\a} \hat \nabla_{\b} \vert =
\half \{ \hat \nabla_{\a}, \hat \nabla_{\b} \} \vert }
while the $\theta^{\a}$ component of $\hat \nabla_{\mu}$ is:
\eqn\vectcomp{ \hat \nabla_{\a} \hat \nabla_{\mu} \vert =
[ \hat \nabla_{\a}, \hat \nabla_{\mu} ] \vert
+ \hat \nabla_{\mu} \hat \nabla_{\a} \vert = }
$$
= [ \hat \nabla_{\a}, \hat \nabla_{\mu} ] \vert
+ \DD_{\mu} \hat \nabla_{\a} \vert + \psi_{\mu}^{\b}
\hat \nabla_{\b} \hat \nabla_{\a} \vert + \psi_{\mu}^{\dot \b}
\hat \nabla_{\dot \b} \hat \nabla_{\a} \vert
$$

{}From \spincomp\ and \vsugra\ we obtain the following results:
$$
\hat \nabla_{+} \hat \nabla_{+} \vert = i \la_{+} \p_{+}
\quad , \quad
\hat \nabla_{-} \hat \nabla_{-} \vert = i \la_{-} \p_{-}
$$
$$
\hat \nabla_{\dot +} \hat \nabla_{\dot +} \vert =
- i \la_{\dot +} \p_{\dot +}
\quad , \quad
\hat \nabla_{\dot -} \hat \nabla_{\dot -} \vert =
- i \la_{\dot -} \p_{\dot -}
$$
$$
\hat \nabla_{+} \hat \nabla_{-} \vert =
- {1 \over 4} \bar H \CX
\quad , \quad
\hat \nabla_{+} \hat \nabla_{\dot -} \vert =
- {1 \over 4} \bar G \CX
$$
$$
\hat \nabla_{\dot +} \hat \nabla_{\dot -} \vert =
- {1 \over 4} H \CX
\quad , \quad
\hat \nabla_{\dot +} \hat \nabla_{-} \vert =
- {1 \over 4} G \CX
$$
$$
\hat \nabla_{+} \hat \nabla_{\dot +} \vert =
{i \over 2} \DD_{\pp}
+ {i \over 2} (\psi_{\pp}^{+} + \la_{\dot +}) \p_{+}
+ {i \over 2} \psi_{\pp}^{-} \p_{-}
+ {i \over 2} (\psi_{\pp}^{\dot +} - \la_{+}) \p_{\dot +}
+ {i \over 2} \psi_{\pp}^{\dot -} \p_{\dot -}
$$
$$
\hat \nabla_{-} \hat \nabla_{\dot -} \vert =
{i \over 2} \DD_{\mm}
+ {i \over 2} \psi_{\mm}^{+} \p_{+}
+ {i \over 2} (\psi_{\mm}^{-} + \la_{\dot -}) \p_{-}
+ {i \over 2} \psi_{\mm}^{\dot +} \p_{\dot +}
+ {i \over 2} (\psi_{\mm}^{\dot -} - \la_{-}) \p_{\dot -}
$$
and from \vectcomp\ we derive the series of identities
that appears below:
$$
\hat \nabla_{+} \hat \nabla_{\pp} \vert =
\DD_{\pp} \hat \nabla_{+} \vert
+ {i \over 2} \psi_{\pp}^{\dot +} \DD_{\pp} \vert
- {1 \over 4} (\psi_{\pp}^{-} \bar H
+ \psi_{\pp}^{\dot -} \bar G) \CX
+ \Big[ i \psi_{\pp}^{+} \la_{+}
+ {i \over 2} \psi_{\pp}^{\dot +}
(\psi_{\pp}^{+} + \la_{\dot +}) -
$$
$$
- 2i \la_{+} \la_{\dot +} -
(\hat \nabla_{\dot +} \la_{+})
- (\hat \nabla_{+} \la_{\dot +}) \Big] \p_{+}
+ {i \over 2} \psi_{\pp}^{\dot +} \psi_{\pp}^{-} \p_{-}
+ {i \over 2} \psi_{\pp}^{\dot +}
(\psi_{\pp}^{\dot +} - \la_{+}) \p_{\dot +}
+ {i \over 2} \psi_{\pp}^{\dot +} \psi_{\pp}^{\dot -} \p_{\dot -}
$$
$$
\hat \nabla_{-} \hat \nabla_{\pp} \vert =
\DD_{\pp} \hat \nabla_{-} \vert
+ {i \over 2} \psi_{\mm}^{\dot -} \DD_{\mm} \vert
- \Big[ {i \over 2} (\hat \nabla_{\dot +} \bar R)
+ {i \over 2} (\hat \nabla_{+} F) + \half (\la_{\dot +}
+ \half \psi_{\pp}^{+}) H - \half (\la_{+}
- \half \psi_{\pp}^{\dot +}) G \Big] \CX +
$$
$$
+ {i \over 2} (G + \psi_{\mm}^{\dot -}
\psi_{\mm}^{+}) \p_{+}
+ {i \over 2} (2 \psi_{\pp}^{-} \la_{-}
+ \psi_{\mm}^{\dot -} (\psi_{\mm}^{-}
+ \la_{\dot -})) \p_{-} + {i \over 2} (\bar H
+ \psi_{\mm}^{\dot -} \psi_{\mm}^{\dot +}) \p_{\dot +}
+ {i \over 2} \psi_{\mm}^{\dot -} (\psi_{\mm}^{\dot -}
- \la_{-}) \p_{\dot -}
$$
$$
\hat \nabla_{\dot +} \hat \nabla_{\pp} \vert =
\DD_{\pp} \hat \nabla_{\dot +} \vert
+ {i \over 2} \psi_{\pp}^{\dot +} \DD_{\pp} \vert
- {1 \over 4} (\psi_{\pp}^{-} G + \psi_{\pp}^{\dot -} H) \CX
+ {i \over 2} \psi_{\pp}^{+} (\psi_{\pp}^{+}
+ \la_{\dot +}) \p_{+}
+ {i \over 2} \psi_{\pp}^{+} \psi_{\pp}^{-} \p_{-} +
$$
$$
+ \Big[ 2i \la_{+} \la_{\dot +}
+ (\hat \nabla_{\dot +} \la_{+})
+ (\hat \nabla_{+} \la_{\dot +})
+ {i \over 2} \psi_{\pp}^{+} (\psi_{\pp}^{\dot +} - \la_{+})
- i \psi_{\pp}^{\dot +} \la_{\dot +} \Big] \p_{\dot +}
+ {i \over 2} \psi_{\pp}^{+} \psi_{\pp}^{\dot -} \p_{\dot -}
$$
$$
\hat \nabla_{\dot -} \hat \nabla_{\pp} \vert =
\DD_{\pp} \hat \nabla_{\dot -} \vert
+ {i \over 2} \psi_{\pp}^{-} \DD_{\mm} \vert
+ \Big[ {i \over 2} (\hat \nabla_{+} R)
+ {i \over 2} (\hat \nabla_{\dot +} \bar F)
- \half \la_{+} H + \half \la_{\dot +} \bar G
- {1 \over 4} \psi_{\pp}^{+} \bar G
- {1 \over 4} \psi_{\pp}^{\dot +} \bar H \Big] \CX +
$$
$$
+ {i \over 2} (\psi_{\pp}^{-} \psi_{\mm}^{+}
- H) \p_{+} + {i \over 2} \psi_{\pp}^{-}
(\psi_{\mm}^{-} + \la_{\dot -}) \p_{-}
+ {i \over 2} (\psi_{\pp}^{-} \psi_{\mm}^{\dot +}
- \bar G) \p_{\dot +}
+ {i \over 2} (\psi_{\pp}^{-} \psi_{\mm}^{\dot -}
- \psi_{\pp}^{-} \la_{-}
- 2 \psi_{\pp}^{\dot -} \la_{\dot -} ) \p_{\dot -}
$$
$$
\hat \nabla_{+} \hat \nabla_{\mm} \vert =
\DD_{\mm} \hat \nabla_{+} \vert
+ {i \over 2} \psi_{\mm}^{\dot -} \DD_{\mm} \vert
- \Big[ {i \over 2} (\hat \nabla_{\dot -} \bar R)
+ {i \over 2} (\hat \nabla_{-} \bar F)
+ \half \la_{\dot -} \bar H - \half \la_{-} \bar G
+ {1 \over 4} \psi_{\mm}^{+} \bar H
+ {1 \over 4} \psi_{\mm}^{\dot +} G \Big] \CX +
$$
$$
+ {i \over 2} \psi_{\mm}^{\dot -} \psi_{\mm}^{+} \p_{+}
+ {i \over 2} ( 2 \psi_{\mm}^{-} \la_{-} - \bar G
+ \psi_{\mm}^{\dot -} (\psi_{\mm}^{-}
+ \la_{\dot -})) \p_{-}
+ {i \over 2} \psi_{\mm}^{\dot -}
\psi_{\mm}^{\dot +} \p_{\dot +}
+ {i \over 2} (\psi_{\mm}^{\dot -}
(\psi_{\mm}^{\dot -} - \la_{-}) - \bar H) \p_{\dot -}
$$
$$
\hat \nabla_{-} \hat \nabla_{\mm} \vert =
\DD_{\mm} \hat \nabla_{-} \vert
+ {i \over 2} \psi_{\mm}^{\dot -} \DD_{\mm} \vert
- {1 \over 4} (\psi_{\mm}^{+} \bar H
+ \psi_{\mm}^{\dot +} G) \CX
+ {i \over 2} \psi_{\mm}^{\dot -} \psi_{\mm}^{+} \p_{+}
+ \Big[ i \psi_{\mm}^{-} \la_{-} +
$$
$$
+ {i \over 2} \psi_{\mm}^{\dot -}
(\psi_{\mm}^{-} + \la_{\dot -}) - 2i \la_{-} \la_{\dot -}
- (\hat \nabla_{\dot -} \la_{-})
- (\hat \nabla_{-} \la_{\dot -}) \Big] \p_{-}
+ {i \over 2} \psi_{\mm}^{\dot -}
\psi_{\mm}^{\dot +} \p_{\dot +}
+ {i \over 2} \psi_{\mm}^{\dot -}
(\psi_{\mm}^{\dot -} - \la_{-}) \p_{\dot -}
$$
$$
\hat \nabla_{\dot +} \hat \nabla_{\mm} \vert =
\DD_{\mm} \hat \nabla_{\dot +} \vert
+ {i \over 2} \psi_{\mm}^{+} \DD_{\pp} \vert
+ \Big[ {i \over 2} (\hat \nabla_{-} R)
+ {i \over 2} (\hat \nabla_{\dot -} F)
- \half \la_{-} H + \half \la_{\dot -} G
- {1 \over 4} \psi_{\mm}^{-} G
- {1 \over 4} \psi_{\mm}^{\dot -} H \Big] \CX +
$$
$$
+ {i \over 2} \psi_{\mm}^{+} (\psi_{\pp}^{+}
+ \la_{\dot +}) \p_{+} + {i \over 2}
(H + \psi_{\mm}^{+} \psi_{\pp}^{-}) \p_{-}
+ {i \over 2} (\psi_{\mm}^{+}
(\psi_{\pp}^{\dot +} - \la_{+})
- 2 \psi_{\mm}^{\dot +} \la_{\dot +}) \p_{\dot +}
+ {i \over 2} (G + \psi_{\mm}^{+}
\psi_{\pp}^{\dot -}) \p_{\dot -}
$$
$$
\hat \nabla_{\dot -} \hat \nabla_{\mm} \vert =
\DD_{\mm} \hat \nabla_{\dot -} \vert
+ {i \over 2} \psi_{\mm}^{-} \DD_{\mm} \vert
- {1 \over 4} (\psi_{\mm}^{+} \bar G
+ \psi_{\mm}^{\dot +} H) \CX
+ {i \over 2} \psi_{\mm}^{-} \psi_{\mm}^{+} \p_{+}
+ {i \over 2} \psi_{\mm}^{-} (\psi_{\mm}^{-}
+ \la_{\dot -}) \p_{-} +
$$
$$
+ {i \over 2} \psi_{\mm}^{-}
\psi_{\mm}^{\dot +} \p_{\dot +} +
+ \Big[ 2i \la_{-} \la_{\dot -}
+ (\hat \nabla_{\dot -} \la_{-})
+ (\hat \nabla_{-} \la_{\dot -})
+ {i \over 2} \psi_{\mm}^{-}
(\psi_{\mm}^{\dot -} - \la_{-})
- i \psi_{\mm}^{\dot -} \la_{\dot -} \Big] \p_{\dot -}
$$
where we also used the commutation relations \newcommut.

Furthermore, the $\hat \nabla_{\pp}$ component
of $\hat \nabla_{\mm}$ is given by:
$$
\hat \nabla_{\pp} \hat \nabla_{\mm} \vert
= \hat \nabla_{\pp} \vert \hat \nabla_{\mm} \vert
+ \psi_{\pp}^{\a} \hat \nabla_{\a} \hat \nabla_{\mm} \vert
+ \psi_{\pp}^{\dot \a} \hat \nabla_{\dot \a}
\hat \nabla_{\mm} \vert
$$
so that:
$$
[ \hat \nabla_{\pp} , \hat \nabla_{\mm} ] \vert =
[ \hat \nabla_{\pp} \vert , \hat \nabla_{\mm} \vert ]
+ \psi_{\pp}^{\a} \hat \nabla_{\a} \hat \nabla_{\mm} \vert
+ \psi_{\pp}^{\dot \a} \hat \nabla_{\dot \a}
\hat \nabla_{\mm} \vert
- \psi_{\mm}^{\a} \hat \nabla_{\a} \hat \nabla_{\pp} \vert
- \psi_{\mm}^{\dot \a} \hat \nabla_{\dot \a}
\hat \nabla_{\pp} \vert =
$$
$$
= [ \DD_{\pp} , \DD_{\mm} ]
+ \DD_{[\pp} (\psi_{\mm ]}^{\a} \p_{\a})
+ \DD_{[\pp} (\psi_{\mm ]}^{\dot \a} \p_{\dot \a})
+ \psi_{\pp}^{\a} \hat \nabla_{\a} \hat \nabla_{\mm} \vert
+ \psi_{\pp}^{\dot \a} \hat \nabla_{\dot \a}
\hat \nabla_{\mm} \vert
- \psi_{\mm}^{\a} \hat \nabla_{\a} \hat \nabla_{\pp} \vert
- \psi_{\mm}^{\dot \a} \hat \nabla_{\dot \a}
\hat \nabla_{\pp} \vert
$$
Substituting the above expressions for the components
of $\hat \nabla_{\a} \hat \nabla_{\mu} \vert$ into
$[ \hat \nabla_{\pp} , \hat \nabla_{\mm} ] \vert$, we get:
\eqn\superformula{
[ \hat \nabla_{\pp} , \hat \nabla_{\mm} ] \vert =
[ \DD_{\pp} , \DD_{\mm} ]
+ \DD_{[\pp} (\psi_{\mm ]}^{\a} \p_{\a})
+ \DD_{[\pp} (\psi_{\mm ]}^{\dot \a} \p_{\dot \a}) + }
%
$$
+ \psi_{\pp}^{+} \Big(
\DD_{\mm} \hat \nabla_{\dot +} \vert
+ {i \over 2} \psi_{\mm}^{+} \DD_{\pp} \vert
+ \Big[ {i \over 2} (\hat \nabla_{-} R)
+ {i \over 2} (\hat \nabla_{\dot -} F)
- \half \la_{-} H + \half \la_{\dot -} G
- {1 \over 4} \psi_{\mm}^{-} G
- {1 \over 4} \psi_{\mm}^{\dot -} H \Big] \CX +
$$
$$
+ {i \over 2} \psi_{\mm}^{+} (\psi_{\pp}^{+}
+ \la_{\dot +}) \p_{+} + {i \over 2}
(H + \psi_{\mm}^{+} \psi_{\pp}^{-}) \p_{-}
+ {i \over 2} (\psi_{\mm}^{+}
(\psi_{\pp}^{\dot +} - \la_{+})
- 2 \psi_{\mm}^{\dot +} \la_{\dot +}) \p_{\dot +}
+ {i \over 2} (G + \psi_{\mm}^{+}
\psi_{\pp}^{\dot -}) \p_{\dot -} \Big) +
$$
$$
+ \psi_{\pp}^{-} \Big(
\DD_{\mm} \hat \nabla_{-} \vert
+ {i \over 2} \psi_{\mm}^{\dot -} \DD_{\mm} \vert
- {1 \over 4} (\psi_{\mm}^{+} \bar H
+ \psi_{\mm}^{\dot +} G) \CX
+ {i \over 2} \psi_{\mm}^{\dot -} \psi_{\mm}^{+} \p_{+}
+ \Big[ i \psi_{\mm}^{-} \la_{-} +
$$
$$
+ {i \over 2} \psi_{\mm}^{\dot -}
(\psi_{\mm}^{-} + \la_{\dot -}) - 2i \la_{-} \la_{\dot -}
- (\hat \nabla_{\dot -} \la_{-})
- (\hat \nabla_{-} \la_{\dot -}) \Big] \p_{-}
+ {i \over 2} \psi_{\mm}^{\dot -}
\psi_{\mm}^{\dot +} \p_{\dot +}
+ {i \over 2} \psi_{\mm}^{\dot -}
(\psi_{\mm}^{\dot -} - \la_{-}) \p_{\dot -} \Big) +
$$
$$
+ \psi_{\pp}^{\dot +} \Big(
\DD_{\mm} \hat \nabla_{\dot +} \vert
+ {i \over 2} \psi_{\mm}^{+} \DD_{\pp} \vert
+ \Big[ {i \over 2} (\hat \nabla_{-} R)
+ {i \over 2} (\hat \nabla_{\dot -} F)
- \half \la_{-} H + \half \la_{\dot -} G
- {1 \over 4} \psi_{\mm}^{-} G
- {1 \over 4} \psi_{\mm}^{\dot -} H \Big] \CX +
$$
$$
+ {i \over 2} \psi_{\mm}^{+} (\psi_{\pp}^{+}
+ \la_{\dot +}) \p_{+} + {i \over 2}
(H + \psi_{\mm}^{+} \psi_{\pp}^{-}) \p_{-}
+ {i \over 2} (\psi_{\mm}^{+}
(\psi_{\pp}^{\dot +} - \la_{+})
- 2 \psi_{\mm}^{\dot +} \la_{\dot +}) \p_{\dot +}
+ {i \over 2} (G + \psi_{\mm}^{+}
\psi_{\pp}^{\dot -}) \p_{\dot -} \Big) +
$$
$$
+ \psi_{\pp}^{\dot -} \Big(
\DD_{\mm} \hat \nabla_{\dot -} \vert
+ {i \over 2} \psi_{\mm}^{-} \DD_{\mm} \vert
- {1 \over 4} (\psi_{\mm}^{+} \bar G
+ \psi_{\mm}^{\dot +} H) \CX
+ {i \over 2} \psi_{\mm}^{-} \psi_{\mm}^{+} \p_{+}
+ {i \over 2} \psi_{\mm}^{-} (\psi_{\mm}^{-}
+ \la_{\dot -}) \p_{-} +
$$
$$
+ {i \over 2} \psi_{\mm}^{-}
\psi_{\mm}^{\dot +} \p_{\dot +}
+ \Big[ 2i \la_{-} \la_{\dot -}
+ (\hat \nabla_{\dot -} \la_{-})
+ (\hat \nabla_{-} \la_{\dot -})
+ {i \over 2} \psi_{\mm}^{-}
(\psi_{\mm}^{\dot -} - \la_{-})
- i \psi_{\mm}^{\dot -} \la_{\dot -} \Big] \p_{\dot -} \Big) -
$$
$$
- \psi_{\mm}^{+} \Big(
\DD_{\pp} \hat \nabla_{+} \vert
+ {i \over 2} \psi_{\pp}^{\dot +} \DD_{\pp} \vert
- {1 \over 4} (\psi_{\pp}^{-} \bar H
+ \psi_{\pp}^{\dot -} \bar G) \CX
+ \Big[ i \psi_{\pp}^{+} \la_{+}
+ {i \over 2} \psi_{\pp}^{\dot +}
(\psi_{\pp}^{+} + \la_{\dot +}) -
$$
$$
- 2i \la_{+} \la_{\dot +} -
(\hat \nabla_{\dot +} \la_{+})
- (\hat \nabla_{+} \la_{\dot +}) \Big] \p_{+}
+ {i \over 2} \psi_{\pp}^{\dot +} \psi_{\pp}^{-} \p_{-}
+ {i \over 2} \psi_{\pp}^{\dot +}
(\psi_{\pp}^{\dot +} - \la_{+}) \p_{\dot +}
+ {i \over 2} \psi_{\pp}^{\dot +}
\psi_{\pp}^{\dot -} \p_{\dot -} \Big) -
$$
$$
- \psi_{\mm}^{-} \Big(
\DD_{\pp} \hat \nabla_{-} \vert
+ {i \over 2} \psi_{\mm}^{\dot -} \DD_{\mm} \vert
- \Big[ {i \over 2} (\hat \nabla_{\dot +} \bar R)
+ {i \over 2} (\hat \nabla_{+} F) + \half (\la_{\dot +}
+ \half \psi_{\pp}^{+}) H - \half (\la_{+}
- \half \psi_{\pp}^{\dot +}) G \Big] \CX +
$$
$$
+ {i \over 2} (G + \psi_{\mm}^{\dot -}
\psi_{\mm}^{+}) \p_{+}
+ {i \over 2} (2 \psi_{\pp}^{-} \la_{-}
+ \psi_{\mm}^{\dot -} (\psi_{\mm}^{-}
+ \la_{\dot -})) \p_{-} + {i \over 2} (\bar H
+ \psi_{\mm}^{\dot -} \psi_{\mm}^{\dot +}) \p_{\dot +}
+ {i \over 2} \psi_{\mm}^{\dot -} (\psi_{\mm}^{\dot -}
- \la_{-}) \p_{\dot -} \Big) -
$$
$$
- \psi_{\mm}^{\dot +} \Big(
\DD_{\pp} \hat \nabla_{\dot +} \vert
+ {i \over 2} \psi_{\pp}^{\dot +} \DD_{\pp} \vert
- {1 \over 4} (\psi_{\pp}^{-} G + \psi_{\pp}^{\dot -} H) \CX
+ {i \over 2} \psi_{\pp}^{+} (\psi_{\pp}^{+}
+ \la_{\dot +}) \p_{+}
+ {i \over 2} \psi_{\pp}^{+} \psi_{\pp}^{-} \p_{-} +
$$
$$
+ \Big[ 2i \la_{+} \la_{\dot +}
+ (\hat \nabla_{\dot +} \la_{+})
+ (\hat \nabla_{+} \la_{\dot +})
+ {i \over 2} \psi_{\pp}^{+} (\psi_{\pp}^{\dot +} - \la_{+})
- i \psi_{\pp}^{\dot +} \la_{\dot +} \Big] \p_{\dot +}
+ {i \over 2} \psi_{\pp}^{+}
\psi_{\pp}^{\dot -} \p_{\dot -} \Big) -
$$
$$
- \psi_{\mm}^{\dot -} \Big(
\DD_{\pp} \hat \nabla_{\dot -} \vert
+ {i \over 2} \psi_{\pp}^{-} \DD_{\mm} \vert
+ \Big[ {i \over 2} (\hat \nabla_{+} R)
+ {i \over 2} (\hat \nabla_{\dot +} \bar F)
- \half \la_{+} H + \half \la_{\dot +} \bar G
- {1 \over 4} \psi_{\pp}^{+} \bar G
- {1 \over 4} \psi_{\pp}^{\dot +} \bar H \Big] \CX +
$$
$$
+ {i \over 2} (\psi_{\pp}^{-} \psi_{\mm}^{+}
- H) \p_{+} + {i \over 2} \psi_{\pp}^{-}
(\psi_{\mm}^{-} + \la_{\dot -}) \p_{-}
+ {i \over 2} (\psi_{\pp}^{-} \psi_{\mm}^{\dot +}
- \bar G) \p_{\dot +}
+ {i \over 2} (\psi_{\pp}^{-} \psi_{\mm}^{\dot -}
- \psi_{\pp}^{-} \la_{-}
- 2 \psi_{\pp}^{\dot -} \la_{\dot -} ) \p_{\dot -} \Big)
$$

Now one can compare this huge formula with the leading
component of the commutator
$[ \hat \nabla_{\pp} , \hat \nabla_{\mm} ] $
computed directly from the Bianchi identities \jacobi:
\eqn\ppmmcomm{
[ \hat \nabla_{\pp} , \hat \nabla_{\mm} ] =
\Big( \la_{\dot +} S - \la_{+} \bar S +
i (\hat \nabla_{\dot +} S) + i (\hat \nabla_{+} \bar S) \Big) \CX +}
$$
+ \Big( {i \over 2} \la_{\dot -} \bar R
- {i \over 2} \la_{-} \bar F
- \half (\hat \nabla_{\dot -} \bar R)
- \half (\hat \nabla_{-} \bar F)
\Big) \hat \nabla_{\dot +}
+ \Big( {i \over 2} \la_{-} R
- {i \over 2} \la_{\dot -} F
+ \half (\hat \nabla_{-} \bar R)
+ \half (\hat \nabla_{\dot -} F)
\Big) \hat \nabla_{+} +
$$
$$
+ \Big( {i \over 2} \la_{\dot +} \bar R
+ {i \over 2} \la_{+} F
- \half (\hat \nabla_{\dot +} \bar R)
+ \half (\hat \nabla_{+} F)
\Big) \hat \nabla_{\dot -}
+ \Big( {i \over 2} \la_{\dot +} \bar F
+ {i \over 2} \la_{+} R
- \half (\hat \nabla_{\dot +} \bar F)
+ \half (\hat \nabla_{+} R)
\Big) \hat \nabla_{-}
$$
where we denoted:
$$
S = {i \over 2} (\hat \nabla_{\dot -} \bar R)
+ {i \over 2} (\hat \nabla_{-} \bar F)
+ \half \la_{\dot -} \bar R
- \half \la_{-} \bar F
$$

By comparing the coefficients of $\hat \nabla_{+} \vert = \p_{+}$
in \superformula\ and \ppmmcomm, for example, we find:
$$
{i \over 2} \la_{-} R - {i \over 2} \la_{\dot -} F
+ \half (\hat \nabla_{-} \bar R)
+ \half (\hat \nabla_{\dot -} F)
= \DD_{[\pp} \psi_{\mm ]}^{+}
+ {i \over 2} \psi_{\pp}^{+} \psi_{\mm}^{+}
(\psi_{\pp}^{+} + \la_{\dot +}) +
$$
$$
+ {i \over 2} \psi_{\pp}^{-} \psi_{\mm}^{\dot -} \psi_{\mm}^{+}
+ {i \over 2} \psi_{\pp}^{\dot +}
\psi_{\mm}^{+} (\psi_{\pp}^{+} + \la_{\dot +})
+ {i \over 2} \psi_{\pp}^{\dot -}
\psi_{\mm}^{-} \psi_{\mm}^{+}
- \psi_{\mm}^{+} \Big[ i \psi_{\pp}^{+} \la_{+}
+ {i \over 2} \psi_{\pp}^{\dot +}
(\psi_{\pp}^{+} + \la_{\dot +})
- 2i \la_{+} \la_{\dot +} -
$$
$$
- (\hat \nabla_{\dot +} \la_{+})
- (\hat \nabla_{+} \la_{\dot +}) \Big]
- {i \over 2} \psi_{\mm}^{-} (G + \psi_{\mm}^{\dot -}
\psi_{\mm}^{+})
- {i \over 2} \psi_{\mm}^{\dot +} \psi_{\pp}^{+}
(\psi_{\pp}^{+} + \la_{\dot +})
- {i \over 2} \psi_{\mm}^{\dot -}
(\psi_{\pp}^{-} \psi_{\mm}^{+} - H)
$$
In the same fashion we obtain the relations defining the other
components of the superfields $R$ and $F$:
$$
{i \over 2} \la_{\dot +} \bar F
+ {i \over 2} \la_{+} R
- \half (\hat \nabla_{\dot +} \bar F)
+ \half (\hat \nabla_{+} R)
= \DD_{[\pp} \psi_{\mm ]}^{-}
+ {i \over 2} \psi_{\pp}^{+}
(H + \psi_{\mm}^{+} \psi_{\pp}^{-})
+ \psi_{\pp}^{-} \Big[ i \psi_{\mm}^{-} \la_{-} +
$$
$$
+ {i \over 2} \psi_{\mm}^{\dot -}
(\psi_{\mm}^{-} + \la_{\dot -}) - 2i \la_{-} \la_{\dot -}
- (\hat \nabla_{\dot -} \la_{-})
- (\hat \nabla_{-} \la_{\dot -}) \Big]
+ {i \over 2} \psi_{\pp}^{\dot +}
(H + \psi_{\mm}^{+} \psi_{\pp}^{-})
+ {i \over 2} \psi_{\pp}^{\dot -} \psi_{\mm}^{-}
(\psi_{\mm}^{-} + \la_{\dot -}) -
$$
$$
- {i \over 2} \psi_{\mm}^{+}
\psi_{\pp}^{\dot +} \psi_{\pp}^{-}
- {i \over 2} \psi_{\mm}^{-} (2 \psi_{\pp}^{-} \la_{-}
+ \psi_{\mm}^{\dot -} (\psi_{\mm}^{-} + \la_{\dot -}))
- {i \over 2} \psi_{\mm}^{\dot +}
\psi_{\pp}^{+} \psi_{\pp}^{-}
- {i \over 2} \psi_{\mm}^{\dot -} \psi_{\pp}^{-}
(\psi_{\mm}^{-} + \la_{\dot -})
$$
$$
{i \over 2} \la_{\dot -} \bar R
- {i \over 2} \la_{-} \bar F
- \half (\hat \nabla_{\dot -} \bar R)
- \half (\hat \nabla_{-} \bar F) =
$$
$$
= \DD_{[\pp} \psi_{\mm ]}^{\dot +}
+ {i \over 2} \psi_{\pp}^{+} (\psi_{\mm}^{+}
(\psi_{\pp}^{\dot +} - \la_{+})
- 2 \psi_{\mm}^{\dot +} \la_{\dot +})
+ {i \over 2} \psi_{\pp}^{-} \psi_{\mm}^{\dot -} \psi_{\mm}^{\dot +}
+ {i \over 2} \psi_{\pp}^{\dot +} (\psi_{\mm}^{+}
(\psi_{\pp}^{\dot +} - \la_{+}) - 2 \psi_{\mm}^{\dot +} \la_{\dot +}) +
$$
$$
+ {i \over 2} \psi_{\pp}^{\dot -} \psi_{\mm}^{-} \psi_{\mm}^{\dot +}
- {i \over 2} \psi_{\mm}^{+} \psi_{\pp}^{\dot +}
(\psi_{\pp}^{\dot +} - \la_{+})
- {i \over 2} \psi_{\mm}^{-} (\bar H
+ \psi_{\mm}^{\dot -} \psi_{\mm}^{\dot +})
- \psi_{\mm}^{\dot +} \Big[ 2i \la_{+} \la_{\dot +}
+ (\hat \nabla_{\dot +} \la_{+}) +
$$
$$
+ (\hat \nabla_{+} \la_{\dot +})
+ {i \over 2} \psi_{\pp}^{+} (\psi_{\pp}^{\dot +} - \la_{+})
- i \psi_{\pp}^{\dot +} \la_{\dot +} \Big]
- {i \over 2} \psi_{\mm}^{\dot -}
(\psi_{\pp}^{-} \psi_{\mm}^{\dot +} - \bar G)
$$
$$
{i \over 2} \la_{\dot +} \bar R + {i \over 2} \la_{+} F
- \half (\hat \nabla_{\dot +} \bar R) + \half (\hat \nabla_{+} F) =
$$
$$
= \DD_{[\pp} \psi_{\mm ]}^{\dot -}
+ {i \over 2} \psi_{\pp}^{+} (G + \psi_{\mm}^{+} \psi_{\pp}^{\dot -})
+ {i \over 2} \psi_{\pp}^{-} \psi_{\mm}^{\dot -}
(\psi_{\mm}^{\dot -} - \la_{-})
+ {i \over 2} \psi_{\pp}^{\dot +} (G + \psi_{\mm}^{+} \psi_{\pp}^{\dot -}) +
$$
$$
+ \psi_{\pp}^{\dot -} \Big[ 2i \la_{-} \la_{\dot -}
+ (\hat \nabla_{\dot -} \la_{-}) + (\hat \nabla_{-} \la_{\dot -})
+ {i \over 2} \psi_{\mm}^{-} (\psi_{\mm}^{\dot -} - \la_{-})
- i \psi_{\mm}^{\dot -} \la_{\dot -} \Big]
- {i \over 2} \psi_{\mm}^{+} \psi_{\pp}^{\dot +} \psi_{\pp}^{\dot -} -
$$
$$
- {i \over 2} \psi_{\mm}^{-} \psi_{\mm}^{\dot -}
(\psi_{\mm}^{\dot -} - \la_{-})
- {i \over 2} \psi_{\mm}^{\dot +} \psi_{\pp}^{+} \psi_{\pp}^{\dot -}
- {i \over 2} \psi_{\mm}^{\dot -} (\psi_{\pp}^{-} \psi_{\mm}^{\dot -}
- \psi_{\pp}^{-} \la_{-} - 2 \psi_{\pp}^{\dot -} \la_{\dot -} )
$$
$$
\la_{\dot +} S - \la_{+} \bar S +
i (\hat \nabla_{\dot +} S) + i (\hat \nabla_{+} \bar S) =
$$
$$
= [ \DD_{\pp} , \DD_{\mm} ]_{\CX}
+ \psi_{\pp}^{+} \Big[ {i \over 2} (\hat \nabla_{-} R)
+ {i \over 2} (\hat \nabla_{\dot -} F)
- \half \la_{-} H + \half \la_{\dot -} G
- {1 \over 4} \psi_{\mm}^{-} G
- {1 \over 4} \psi_{\mm}^{\dot -} H \Big]
- {1 \over 4} \psi_{\pp}^{-} (\psi_{\mm}^{+} \bar H +
$$
$$
+ \psi_{\mm}^{\dot +} G)
+ \psi_{\pp}^{\dot +} \Big[ {i \over 2} (\hat \nabla_{-} R)
+ {i \over 2} (\hat \nabla_{\dot -} F) - \half \la_{-} H
+ \half \la_{\dot -} G - {1 \over 4} \psi_{\mm}^{-} G
- {1 \over 4} \psi_{\mm}^{\dot -} H \Big]
- {1 \over 4} \psi_{\pp}^{\dot -}
(\psi_{\mm}^{+} \bar G + \psi_{\mm}^{\dot +} H) +
$$
$$
+ {1 \over 4} \psi_{\mm}^{+} (\psi_{\pp}^{-} \bar H
+ \psi_{\pp}^{\dot -} \bar G)
+ \psi_{\mm}^{-} \Big[ {i \over 2} (\hat \nabla_{\dot +} \bar R)
+ {i \over 2} (\hat \nabla_{+} F) + \half (\la_{\dot +}
+ \half \psi_{\pp}^{+}) H
- \half (\la_{+} - \half \psi_{\pp}^{\dot +}) G \Big] +
$$
$$
+ {1 \over 4} \psi_{\mm}^{\dot +}
(\psi_{\pp}^{-} G + \psi_{\pp}^{\dot -} H)
- \psi_{\mm}^{\dot -} \Big[ {i \over 2} (\hat \nabla_{+} R)
+ {i \over 2} (\hat \nabla_{\dot +} \bar F)
- \half \la_{+} H + \half \la_{\dot +} \bar G
- {1 \over 4} \psi_{\pp}^{+} \bar G
- {1 \over 4} \psi_{\pp}^{\dot +} \bar H \Big]
$$

\appendix{D}{Derivation of the projection formula
in $\CN=(2,2)$ dilaton supergravity}

Here we repeat the derivation \GWmeasure\ of the local
density projection formula in $\CN=(2,2)$ dilaton supergravity.
Namely, we start by writing the most general expression
for the chiral projector with the right dimension and
index structure:
\eqn\cprojtobefound{
\int d^2x d^4 \th E^{-1} \CL = \int d^2x e^{-1} \big[
\hat \nabla_{+} \hat \nabla_{-}
+ X^{+} \hat \nabla_{+} + X^{-} \hat \nabla_{-}
+ Y \big] \hat \nb^2 \CL \vert }
where the coefficients $X^{\a}$ and $Y$ are to be determined.
Following \GWmeasure, we evaluate \cprojtobefound\ for the kinetic
action $\CL = \bar \Phi \Phi$ of a free chiral multiplet:
$$
\int d^2x d^4 \th E^{-1} \bar \Phi \Phi =
\int d^2x e^{-1} \big[
(\hat \nabla^2 \hat \nb^2 \bar \Phi) \Phi \vert
+ (\hat \nabla_{+} \hat \nb^2 \bar \Phi)
(\hat \nabla_{-} \Phi) \vert
-  (\hat \nabla_{-} \hat \nb^2 \bar \Phi)
(\hat \nabla_{+} \Phi) \vert +
$$
\eqn\cprojfree{
+ (\hat \nb^2 \bar \Phi) (\hat \nabla^2 \Phi) \vert
+ X^{+} (\hat \nabla_{+} \hat \nb^2 \bar \Phi) \Phi \vert
+ X^{+} (\hat \nb^2 \bar \Phi) (\hat \nabla_{+} \Phi) \vert }
$$
+ X^{-} (\hat \nabla_{-} \hat \nb^2 \bar \Phi) \Phi \vert
+ X^{-} (\hat \nb^2 \bar \Phi) (\hat \nabla_{-} \Phi) \vert
+ Y (\hat \nb^2 \bar \Phi) \Phi \vert \big]
$$
Clearly, the left-hand side of this formula
is invariant under complex conjugation.
So, the right-hand side must be invariant as well.
As we will see in a moment, this condition completely
determines the unknown coefficients $X^{\a}$ and $Y$.
It suffices to consider only bosonic terms.
Using the formulas in appendix C along with
the definition \cmult, one can easily compute
the relevant components:
$$
\hat \nabla_{+} \hat \nabla_{\dot +}
\hat \nabla_{\dot -} \bar \Phi \vert \sim
- \psi_{\pp}^{-} \DD_{\mm} \bar \phi
+ (\psi_{\pp}^{\dot +} - \la_{+}) \bar A
$$
$$
\hat \nabla_{-} \hat \nabla_{\dot +}
\hat \nabla_{\dot -} \bar \Phi \vert \sim
\psi_{\mm}^{+} \DD_{\pp} \bar \phi
+ (\psi_{\mm}^{\dot -} - \la_{-}) \bar A
$$
$$
\hat \nabla_{+} \hat \nabla_{-} \hat \nabla_{\dot +}
\hat \nabla_{\dot -} \bar \Phi \vert \sim
\DD_{\mm} \DD_{\pp} \bar \phi
+ i (\psi_{\mm}^{\dot +} \psi_{\pp}^{+}
- \la_{+} \psi_{\mm}^{+}) \DD_{\pp} \bar \phi +
$$
$$
+ i (\psi_{\mm}^{\dot -} - \la_{-})
\psi_{\pp}^{-} \DD_{\mm} \bar \phi
- {i \over 4} \bar H \bar A
+ i \psi_{\mm}^{\dot +} \psi_{\pp}^{\dot -} \bar A
- i (\psi_{\mm}^{\dot -} - \la_{-})
(\psi_{\pp}^{\dot +} - \la_{+}) \bar A
$$
Substituting these into \cprojfree, we find that the resulting
component action is invariant under complex conjugation
only if we put:
$$
X^{+} = i (\psi_{\mm}^{\dot -} - \la_{-})
$$
$$
X^{-} = - i (\psi_{\pp}^{\dot +} - \la_{+})
$$
$$
Y = - {1 \over 4} \bar H
- \psi_{\pp}^{\dot -} \psi_{\mm}^{\dot +}
+ (\psi_{\mm}^{\dot -} - \la_{-})
(\psi_{\pp}^{\dot +} - \la_{+})
$$
With these expressions for $X^{\a}$ and $Y$ the result does not
depend on whether we use the chiral projector \cprojtobefound\
or its complex conjugate, of course, as it should be.
Specifically, the chiral projection formula in $\CN=(2,2)$ dilaton
supergravity takes the following form:
\eqn\cprojfound{
\int d^2x d^4 \th E^{-1} \CL = \int d^2x e^{-1} \big[
\hat \nabla_{+} \hat \nabla_{-}
+ i (\psi_{\mm}^{\dot -} - \la_{-}) \hat \nabla_{+}
- i (\psi_{\pp}^{\dot +} - \la_{+}) \hat \nabla_{-} + }
$$
+ \big( - {1 \over 4} \bar H
- \psi_{\pp}^{\dot -} \psi_{\mm}^{\dot +}
+ (\psi_{\mm}^{\dot -} - \la_{-})
(\psi_{\pp}^{\dot +} - \la_{+})
\big) \big] \hat \nb^2 \CL \vert
$$
One can easily check that this expression is equivalent
to the chiral density projector \newcproj\ obtained
by de-gauging the corresponding projector
in the non-minimal $\CN=(2,2)$ supergravity.

\listrefs
\end